\documentclass[a4paper, 10pt, oneside, onecolumn, 3p]{elsarticle}

\usepackage{amssymb}
\usepackage{booktabs}
\usepackage{multirow}
\usepackage{placeins}
\usepackage{framed}
\usepackage{xcolor}
\usepackage{mdframed}
\usepackage{hyperref} 
\usepackage{tcolorbox}
\usepackage{lipsum} 
\usepackage{natbib}
\usepackage{bibentry}
\usepackage{subcaption}
\usepackage{graphicx}
\usepackage{adjustbox}
\usepackage{color}
\usepackage{ragged2e}
\usepackage{enumitem}
\newtcolorbox{myframe}[1][]{
    width=35em,
    top=5pt,
    bottom=5pt,
    left=5pt,
    right=5pt,
    arc=10pt,
    auto outer arc,
    colback=white,
    colframe=black,
    fonttitle=\bfseries,
    title=\centering\textbf{#1}
}

\newcommand{\todo}[1]{}
\renewcommand{\todo}[1]{{\color{red} TODO: {#1}}}

\journal{Information and Software Technology}
\begin{document}

\justify 
\begin{frontmatter}



\title{Requirements Engineering for Older Adult Digital Health Software: A Systematic Literature Review}


\author[inst1]{Yuqing Xiao}  
\author[inst1]{John Grundy}

\affiliation[inst1]{organization={Department of Software Systems and Cybersecurity, Faculty of IT, Monash University},
            city={Clayton},
            postcode={3800}, 
            state={VIC},
            country={Australia  }}

\author[inst2]{Anuradha Madugalla}

\affiliation[inst2]{organization={School of IT, Deakin University},
            city={Melbourne},
            postcode={3125}, 
            state={VIC},
            country={Australia}}

\end{frontmatter}


\section{Introduction}
According to the World Population Prospects 2022, the older adult population of the world is growing rapidly~\cite{un2022world}. The percentage of the global population aged 65 and above is estimated to increase from 10\% in 2022 to 16\% in 2050 ~\cite{un_ageing}. Ageing significantly impacts individual health, leading to crucial challenges of physical impairment and mental decline. Aged care, encompassing support services for older individuals' health and well-being, addresses these challenges and rising needs, while accommodating the diverse physical, emotional, and social aspects of individuals ~\cite{brownie2013effects, grundy2023vision}.

A growing body of literature recognises the importance of software solutions for healthcare problems, including cancer support ~\cite{jiang2020machine}, intensive care monitoring ~\cite{hyland2020early}, and chronic disease tracking ~\cite{subramanian2020precision}. As aged care needs can be viewed as a combination of several healthcare problems, applying software solutions can address this urgent demand. For example, personalized home care can improve the overall quality of care and enhance the well-being of older individuals with fragility and cognitive decline.  Software Engineering for ageing users' digital health software has become a popular area of research and practice.  These tend to focus on improving the quality of care for older adults in home or aged care healthcare settings, including adapting to the changing needs of older adults, developing various ageing user Mobile Health (mHealth) applications, smart home, care solutions, and conducting diverse usability testing with older adults.

Requirements engineering (RE) is a critical part of software engineering. This involves obtaining, analysing, recording, implementing, and updating software requirements. In the development of digital health software for older adults, RE can play a crucial role in ensuring that the software is user-centred, accessible, secure, and relevant to the needs of ageing end users and their caregivers ~\cite{grundy2023vision}. RE is particularly important in medical, health, and welfare software engineering for older end users as they have varying levels of comfort with modern technology, diverse living situations, and diverse physical and mental impacts of ageing.

While several studies have applied RE for their older user-targeted digital health applications, a systematic review has not yet been conducted to identify the key strategies and requirements of diverse software that are developed for ageing users. Several RE frameworks for aged care software solutions have been proposed; however, the analysis of these frameworks remains inadequate to enhance the field's common agreement and mutual learning. Therefore, conducting a systematic literature review (SLR) on requirements engineering for aged care is necessary to ensure that digital health software for older adults meets the specific needs of older adults and is designed to be accessible, secure, and compliant with regulations. The main contributions of this SLR are to 1) highlight the key works that have used RE in older adult digital health; 2) explore how the RE is done in each study; and 3) point out the key benefits, limitations, and recommendations of RE in each study.
In our SLR we analyzed 69 studies that were found by filtering studies from eight databases. Our results reveal that the usage and adoption of requirements in older adult digital health software vary by quality and conditions, which is related to the RE techniques used, requirement modelling, and in which stage RE was involved during system building.

This paper is structured as follows: Section 2 provides background and related work. Section 3 presents the review protocol, details about the conduction, and results reporting. Section 4 discusses the advances so far, and Section 5 presents final remarks. This endeavour can ultimately lead to a more successful app that improves the health and well-being of older adults with chronic conditions.

\section{Related Work}
\label{sec:Related Work}

\subsection{Requirements Engineering}

Requirements Engineering (RE) is essential for developing quality software systems ~\cite{laplante2022requirements}. Key RE-related include identifying stakeholders, eliciting and analyzing requirements, and documenting them for developers. Common RE models include user personas ~\cite{karolita2023use}, user stories ~\cite{dalpiaz2020conceptualizing}, use cases ~\cite{dalpiaz2020conceptualizing}, and various semantic models ~\cite{belani_towards_2022}. 
They help to clarify and specify requirements in a way that is more precise and unambiguous, involving ontology models, conceptual models, and even Natural Language Processing (NLP) models. Heyn et. al has mentioned that using RE conceptual models can improve contextual definitions, data attributes, performance definition and monitoring, and the impact of human factors on system acceptance and success ~\cite{heyn2021requirement}. Nazir et al. and Ahmed et al. have verified that NLP-based and Large Language Model (LLM)s requirements engineering can greatly reduce the amount of rework and improve qualities of the system\cite{nazir2017applications, ahmad2023requirements}. 

In this review, due to the terms used in many studies being diverse, we also take into account the studies that applied user goal modelling, co-design, participatory design process, and user-centred approach, as long as there is a significant RE aspect described in the studies. User Needs Assessment identifies and prioritizes the needs and preferences of users ~\cite{billings1995approaches, watkins1998needs, sleezer2014practical}. Differing from RE, user needs assessment may focus more on understanding user preferences and behaviours rather than specific software requirements. User Goal Modelling defines the specific aims that users want to achieve with a software system. Co-Design is a collaborative design approach that involves users, designers, and other stakeholders working together to design solutions and actively involves users in the design process. Participatory design involves users in the design process, allowing them to contribute ideas, feedback, and suggestions. It encompasses various design activities beyond requirements elicitation, such as prototyping, usability testing, and iterative design. User-centred Approaches are an overarching philosophy that prioritizes the needs, preferences, and capabilities of users throughout the design and development process. 
 

\subsection{Digital Health Softwares for Ageing Users}

Digital health, or eHealth, involves using information and communication technologies to enhance healthcare delivery ~\cite{sharma2018using, thomas2014review, batra2017digital}. 
It includes electronic health records (EHRs), telemedicine, mHealth apps, wearable devices, and health information systems, enabling better communication among healthcare professionals, remote patient monitoring, and empowering individuals with real-time health data. Digital health benefits patients, caregivers, and society by improving outcomes, decision-making, and resource allocation.
Many digital health technologies have been used in the field of older adult health care, such as remote monitoring and telehealth, medication management, fall detection and prevention, cognitive health, and social connectedness. A systematic review has analysed such work with a special focus on smart homes for older adults ~\cite{pal2017smart}. These solutions can improve the quality of life for older adults, and the review study highlights the potential of smart home technologies to enhance their quality of life. Another study about IoT-based systems and applications designed for aged care digital health examined the key technologies used, their applications in healthcare settings, and their impact on the quality of care for senior individuals ~\cite{matayong2023iot}. They found that IoT-based systems have the potential to significantly improve senior healthcare by enabling remote monitoring, personalized care, and early diagnosis of diseases. 

\subsection{Requirements Engineering for Digital Health}
Developing good digital health software requires a deep understanding of healthcare workflows, user needs, and regulatory requirements to ensure these systems' efficacy, safety, and usability. 
The necessity and challenges of using RE in AI-based systems have been noticed by many researchers ~\cite{heyn2021requirement}, including understanding, determining, and specifying. The common challenge can be derived from contextual definitions, data attributes, performance definition and monitoring, and the impact of human factors on system acceptance and success.  
Due to users and carers particular demographics and needs, RE is crucial for ensuring good digital health software for older adults. 
Previous research on RE in healthcare highlights the challenges and best practices for gathering, analysing, and managing requirements for healthcare systems ~\cite{cysneiros2002requirements}. They emphasize the importance of RE in ensuring the successful development and deployment of healthcare systems that meet the needs of stakeholders. A Systematic literature review on RE underscores the importance of human-centred design in digital health ~\cite{levy2023sustaining}. In this work, they present a RE perspective to sustain health, discussing the challenges and opportunities for applying RE principles to address health-related issues. They suggest that RE is vital for developing sustainable health solutions aligned with user needs and preferences. 



\subsection{Reviews on RE and Digital Health Softwares for Ageing People}









Good requirements elicitation is crucial in digital health ~\cite{aziz2016requirement,ariaeinejad2016user}. Several researchers have explored the status of requirement-gathering activities in the co-design or participatory design processes when they develop solutions for the ageing population in digital health. Fischer et al. ~\cite{fischer2020importance} conducted a systematic review on older users involvement in technology design, highlighting the need for inclusivity and tailored solutions. However, their focus on inclusive intervention in technology design is from a gerontologist's perspective and does not deeply investigate key RE processes and techniques. Merkel and Kucharski ~\cite{merkel2019participatory} also examined participatory design in gerontechnology (Technology for ageing population support), emphasizing the value of involving older adults in the design process for more effective and user-friendly technologies but lacking focus on RE or SE issues. Chute et al. explored user requirements for co-managed digital health and care, applying emphasis towards understanding the diverse needs of users in co-managed care settings ~\cite{chute2022user}. They highlighted the significance of user requirements in developing digital health aged care software, especially in integrated care models. However, their focus on co-managed care for comorbidity did not address issues related to prioritizing, documenting, and analysing requirements. Zhang et al. outlined a scoping review protocol focusing on co-design in residential aged care, showing a growing interest in this area ~\cite{zhang2023application}. This study collectively recognizes the benefits of co-design in enhancing technology and care practices for older adults in aged care facilities, indicating a positive trend towards more person-centred approaches. However, their protocols did not consider RE, nor did they examine the software details of residential aged care. 

\section{Research Methodology}
\label{sec: Research Methodology}
This Systematic Literature Review (SLR) was conducted to investigate and synthesize the current research findings on Requirements Engineering (RE) for digital health software focusing on senior adults as the end users.

To guide our SLR protocol development, three guidelines were followed which are stated below. 
\begin{enumerate}
    \item PRISMA ~\cite{page2021prisma}: General systematic review format guidelines 
    \item The guideline for  Software Engineering Evidence-based \cite{kitchenham2015evidence}
    \item PICO ~\cite{richardson1995well} framework: This was originally developed for medicine, but can be now applied to IT as well
\end{enumerate}
\begin{figure} [h]
    \centering
    \includegraphics[width=0.8\textwidth,height=7cm]{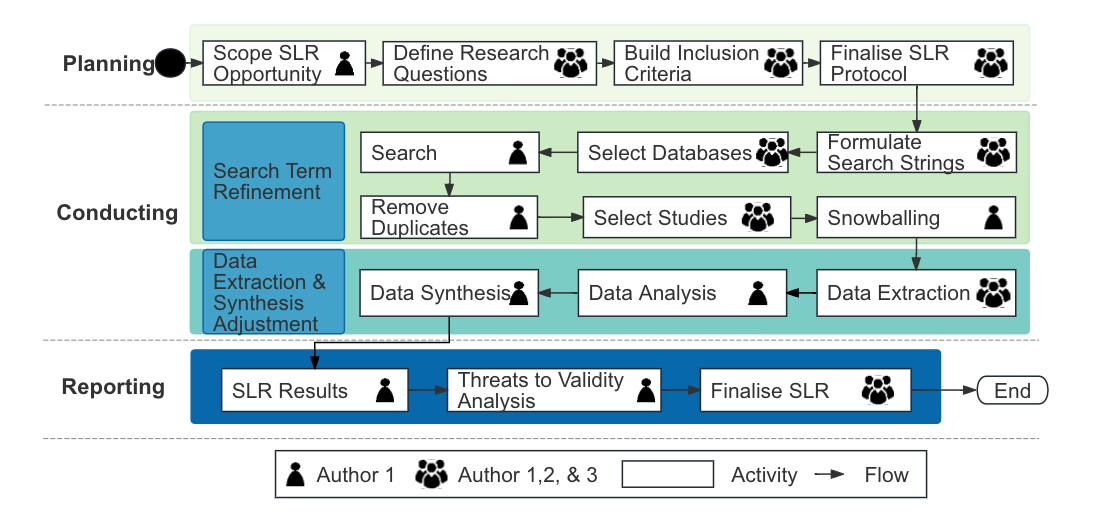}
    \caption{Systematic Literature Review Process}
    \label{fig:SLR_process}
\end{figure}

The process we followed, outlined in the ~\ref{fig:SLR_process}, has three stages: planning, conducting, and reporting. In the planning stage, we identified the need for the SLR, formulated research questions, defined the protocol, and reviewed it. In the conducting stage, we formulated search strings, selected databases, and used the SLR management tool 'Covidence' to remove duplicates and select studies. After snowballing, we ended up with all the studies we wanted and extracted the data. Finally, after the data analysis and synthesis, we reported the SLR results, analysed threats to validity, and finalized the SLR after all authors' reviews. 
\subsection{Research Questions}

We wanted to answer the following key research questions (RQs) in this SLR:
\begin{enumerate}
    \item What are the key research works carried out to date focusing on Requirements Engineering for digital health for ageing people?
        \begin{enumerate}
        \item What is the nature and type of studies carried out?
        \item What aged healthcare and well-being issues are addressed in each work?
        \item What are the demographics of older adult people in each study?
        \item What data is captured by the (proposed) software and how is the data used? What technologies did they use? Does the software use any AI solutions? 
        \item What different human aspects (besides age) were considered in the study (if any)?
         ]
        \end{enumerate}
    
    \item How was the RE carried out?
        \begin{enumerate}
        \item What RE techniques did each study use?
        \item What tools were used for requirements elicitation and documentation in each study?
        \item How were the requirements modelled in each study?
        \item How were the requirements validated?
        \item Were the requirements used to build an actual system? If so, in which SE stages? 
        \item Were the requirements used to evaluate the solution? If so, how is the study evaluated?
        \end{enumerate}
    
    \item What are the key Strengths, Limitations, Gaps, and Future work recommendations in the selected studies?
        \begin{enumerate}
        \item What are the key benefits/positive outcomes reported?
        \item What are the key limitations reported?    
        \item What are key recommendations for future research?

        \end{enumerate}
\end{enumerate}
\subsection{Search Strategy}




We started with a concept map of terms, which can be listed as: Concept 1): Elderly; Older; Ageing; senior; Residential ageing care. Concept 2): Requirement* Engineering; Requirement elicitation; Requirement extraction. Concept 3): Software Engineering; user*; UX; UI; personas; co-design. We used these concepts to formulate our search strings. We searched databases including ACM Digital Library, IEEE Xplore, Inspec, Springer, Wiley, Sage Journals Online, Taylor and Francis Online, and Scopus. 
The use of these databases was motivated by several factors. We incorporated well-known computer science sources such as ACM, IEEE, and Inspec, alongside medical-focused databases like Sage and Taylor \& Francis. 
The key search strings used for different databases are listed in Table\ref{tab:search-string-table}. 

\begin{table}[h!]
\tiny
\centering
\begin{tabular}{p{1.6cm}|p{8cm}}
\hline
ACM      &  \multicolumn{1}{p{10cm}}{ [[Full Text: ``requirement* engineering"] OR [Full Text: ``requirement elicitation"] OR [Full Text: ``requirement extraction"] OR [[Full Text: ``software engineering"] AND [[Full Text: ``ui"] OR [Full Text: ``ux"] OR [Full Text: ``co-design"]]]] AND NOT [Title: ``blockchain"] AND NOT [[Title: ``systematic mapping study"] OR [Title: ``literature review"] OR [Title: systematic review] OR [Title: ``qualitative study"] OR [Title: ``vision paper"]] AND [[Full Text: ``elderly"] OR [Full Text: ``aged care"] OR [Full Text: ``ageing care"] OR [Full Text: ``elderly adult"] OR [Full Text: ``senior adult"] OR [Full Text: ``older adult"] OR [Full Text: ``ageing residential care"]]}   \\ \hline
IEEE      & \multicolumn{1}{p{10cm}}{(``Abstract":``aged care" OR ``Abstract":``elderly adult" OR ``Abstract":``senior adult" OR ``Abstract":``Older adults” OR ``Abstract": ageing OR ``Abstract":Residential ageing care) AND ("All Metadata":``Requirement* Engineering" OR ``requirements extraction" OR ``requirements elicitation" OR (``Software Engineering" AND ``stakeholder" OR UI OR UX OR co-design;) NOT (``All Metadata":``Blockchain")
}  \\ \hline
Inspec & \multicolumn{1}{p{10cm}}{(((``elderly" OR ``aged" OR ``aged care" OR ``ageing care" OR ``elderly adult" OR ``senior adult" OR ``older adult" OR ageing OR ``ageing residential care") WN KY) AND (("Requirement* Engineering" OR ``requirements extraction" OR ``requirements elicitation" OR ("Software Engineering" AND (``stakeholder" OR UI OR UX OR co-design))) WN KY)) NOT ((``blockchain") WN KY)) AND ({english} WN LA))}  \\ \hline
Springer  & \multicolumn{1}{p{10cm}}{("aged care" OR ``ageing care" OR ``elderly adult" OR ``senior adult" OR ``older adult" OR ``aged residential care") AND ("requirement* engineering" OR ``Requirements elicitation" OR ``Requirements extraction" OR ("Software Engineering" AND ( ``stakeholder" OR ``UI" OR ``UX" OR ``co-design" )))' within ``Computer Science" ``Conference Paper" ``English"} \\ \hline
Wiley & \multicolumn{1}{p{10cm}}{````elderly" OR ``aged" OR ``aged care" OR ``ageing care" OR ``elderly adult" OR ``senior adult" OR ``older adult" OR ``aging" OR ``aged residential care"" anywhere and ````requirements engineering" OR ``Requirements elicitation"OR ``Requirements extraction" OR ("Software Engineering" AND ( ``stakeholder" OR ``UI" OR ``UX" OR ``co-design" ))"} \\ \hline
Sage Journals Online & \multicolumn{1}{p{10cm}}{``elderly" OR ``aged" OR ``aged care" OR ``ageing care" OR ``elderly adult" OR ``senior adult" OR ``older adult" OR ``aging" OR ``aged residential care"" anywhere and ``"requirements engineering" OR ``Requirements elicitation"OR ``Requirements extraction" OR (``Software Engineering" AND ( ``stakeholder" OR ``UI" OR ``UX" OR ``co-design" ))} \\ \hline
Taylor and Francis online & \multicolumn{1}{p{10cm}}{[[Abstract: ``elderly"] OR [Abstract: ``aged"] OR [Abstract: ``aged care"] OR [Abstract: ``ageing care"] OR [Abstract: ``elderly adult"] OR [Abstract: ``senior adult"] OR [Abstract: ``older adult"] OR [Abstract: ``aging"] OR [Abstract: ``aged residential care"]] AND [[All: ``requirement elicitation"] OR [All: ``requirement extraction"] OR [All: ``requirement engineering"] OR [All: ``requirements engineering"] OR [[All: ``software engineering"] AND [[All: ``co-design"] OR [All: ``user*"] OR [All: ``ui"] OR [All: ``ux"]]]]}
\\
\hline

\end{tabular}
\caption{Table of Search Strings}
\label{tab:search-string-table}
\end{table}


We constructed a `gold set' of primary studies that comprises a predefined selection of studies serving as a benchmark to evaluate our search strategy. This set was meticulously defined through reviews by all authors after initial searches in key databases like Scopus, ACM, and IEEE, ensuring that the selected articles met the inclusion criteria and were relevant to our research questions. In constructing the search string for each database, our aim was to ensure that the gold set papers were included in the search results. For example, a gold set paper from ACM should be included in the search results when using the ACM search string.
In the event that our initial search results did not encompass all studies within the gold set, we refined our search strategy by revising the search string parameters to include additional studies. 

To address potential oversights in the automated search process, a manual search was conducted in accordance with snowballing guidelines outlined by Wohlin ~\cite{wohlin2014guidelines}. A total of 26 initially selected gold set papers, along with relevant works, were subjected to iterative forward and backward snowballing. This process continued over three iterations until no further pertinent papers were identified. As a result, the snowballing yielded 26 additional papers.

\subsection{Inclusion and Exclusion Criteria}

We defined some key inclusion papers for studies:
\begin{enumerate}
    \item paper is written in English 
    \item aged or ageing users are the main stakeholders/end-users of the software system
    \item focus or a significant part of the study includes RE for ageing people supporting software systems, including in-home, residential care and institutional care
    \item describes a primary study of RE for cognitive or physical support training software
\end{enumerate}

We also defined some exclusion criteria:
\begin{enumerate}
    \item Paper has little or no RE content 
    \item End users or stakeholders are not significantly made up of older adult users, eg. are diverse chronic disease patients, caregivers, clinical staff
    \item Studies that talk about ``aged software systems"
    \item Primary focus is on younger users of a software system
    \item Visionary studies (not a primary study) or secondary studies (SLR, SMS, Survey)
    \item Primary focus is HCI aspects, the focus is only on human-computer interface components, not RE
    \item Short studies, less than 5 pages
\end{enumerate}




\begin{figure}[!ht]
    \centering
    \includegraphics[width=0.8\textwidth]{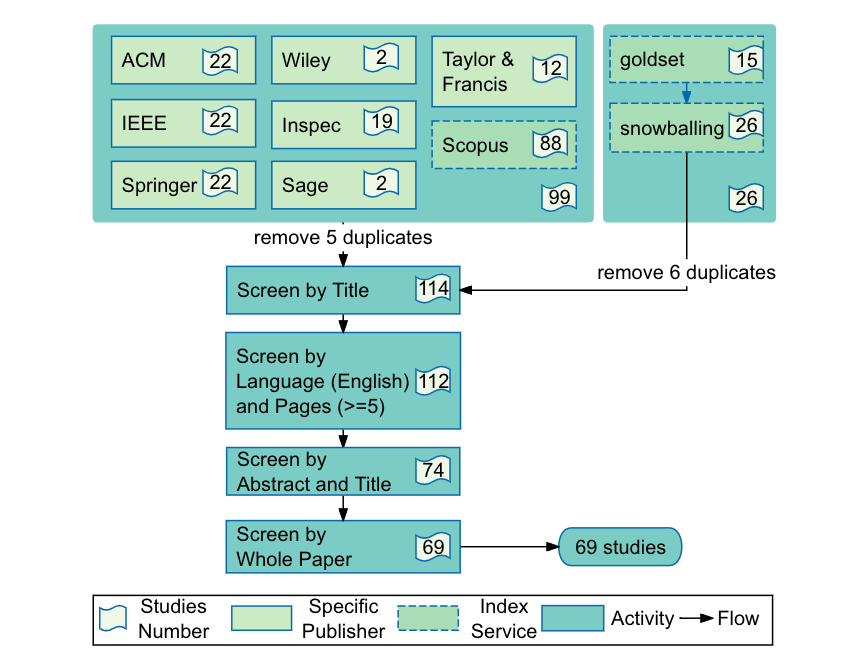}
    \caption{Systematic Literature Review Screen Strategy}
    \label{fig:screen_process}
\end{figure}

\subsection{Selection Process}

The selection process is depicted in Figure \ref{fig:screen_process}. Initially, we identified 99 studies from the 8 databases, reduced to 94 after removing duplicates. Starting with a gold set of 15 papers, the snowballing process added 26 more studies. After combining database and snowballing results, title screening reduced the list to 114 studies. We excluded 2 studies not in English or shorter than 5 pages, leaving 112. Next, we conducted a screening based on abstract and title, proceeding with studies that at least two authors agreed to include, leading to 74 studies. Finally, full paper screening resulted in 69 studies.
We scored the primary studies based on a quality matrix, both outlined in the Online Appendix. We did not use the paper quality score to remove any of the primary studies found.

\section{Results \& Discussion}

\subsection{Study Demographics}
\begin{figure}[!ht]
    \centering
    \includegraphics[width=0.75\textwidth, trim={7cm 0.5cm 7cm 0.5cm},clip]{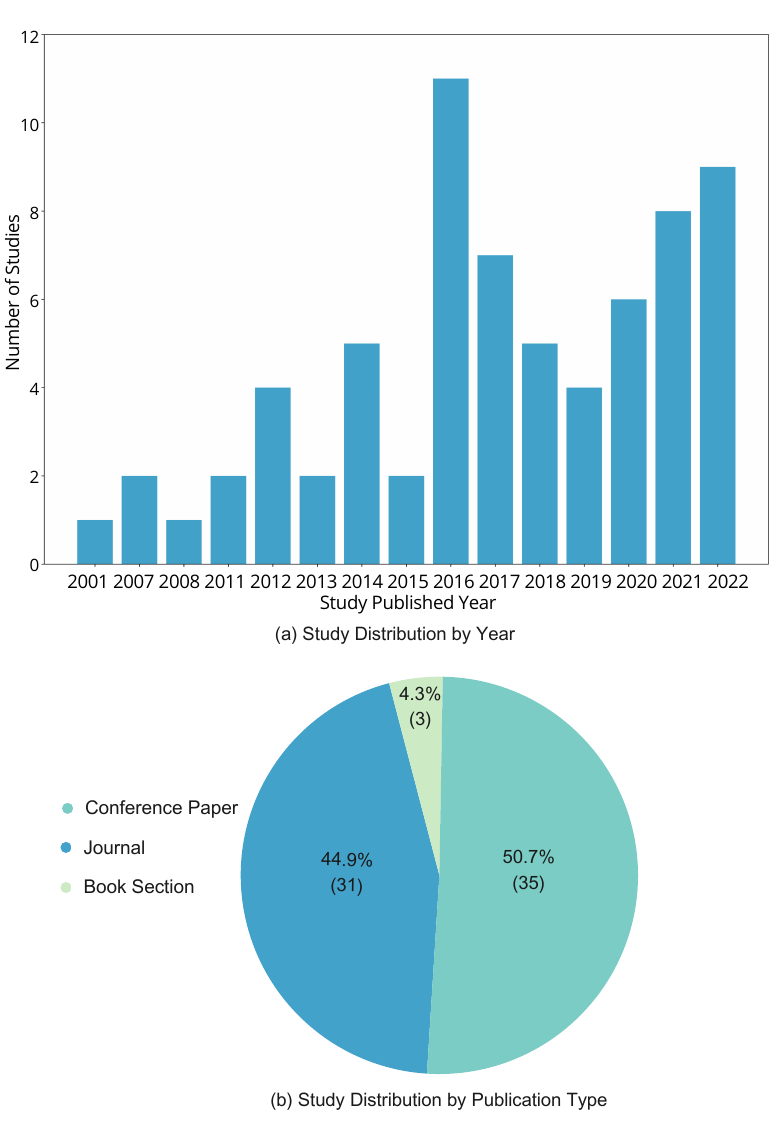}
    \caption{Study Distribution by Year and Type}
    \label{fig:Study Distribution by Year_type}
\end{figure}

After our selection process, we identified 69 primary studies published between 2001 and 2024. Figure \ref{fig:Study Distribution by Year_type}(a) shows a bar graph of study publication years, revealing a significant increase in studies published in 2016 compared to 2015. This might be related to the revolution of Machine learning and AI algorithms around 2016. As some studies have pointed out, after Machine Learning and AI showed their potential in the field of computing, many studies including digital health, HCI, and SE in healthcare have been adopting them ~\cite{mesko2017digital, muller2022genaichi, alshamrani2022iot}. The results suggest rising interest in RE for aged care software systems. 



Figure \ref{fig:Study Distribution by Year_type}(b) illustrates the publication type distribution of our  69 primary studies. The largest accounting for just over half at 50.7\% are conference papers. This is followed by journal articles, which include 31 studies (44.9\%). The remaining 4.3\% are Book Chapters.
Of the total 69 studies, 50 (72.46\%) were from academia and 19 (27.54\%) were from industry (in collaboration with academia). No study was purely industrial-based research.



\begin{table}[h!]
\footnotesize
\centering
\resizebox{0.8\textwidth}{!}{%
\footnotesize
\begin{tabular}{lll}
\textbf{Categories}           & \textbf{Studies}                                           & \textbf{Percentage} \\ \hline
\textbf{No citation}       & S34, S20, S47, S48, S6                           & 7.25\%     \\ \hline
\textbf{1-50} &
  \begin{tabular}[c]{@{}l@{}}S27, S16, S14, S43, S39, S45, S68, S3, S18, \\ S22, S24, S55, S33, S11, S17, S44, S25, S53,\\ S38, S59, S4, S9, S56, S67, S10, S57, S28, \\ S15, S5, S62, S32, S13, S12, S8, S42, S66, \\ S7, S36, S21, S46, S49, S60, S2, S35, S63\end{tabular} &
  65.21\% \\ \hline
\textbf{51-100}               & S19, S64, S69, S29, S23, S31, S50, S52, S65, S41 & 14.49\%    \\ \hline
\textbf{101-200}              & S1, S40, S54, S37, S61, S26                      & 8.70\%     \\ \hline
\textbf{high citation [201+]} & S30, S51, S58                                    & 4.35\%     \\ \hline
\end{tabular}%
}
\caption{Studies Distributed by Citation Numbers}
\label{tab: Studies Distributed by Citation Numbers}
\end{table}

Table \ref{tab: Studies Distributed by Citation Numbers} shows a summary of our selected primary studies based on their citation numbers. We categorized our studies into five groups: no citations (0), 1-50 citations, 51-100 citations, 101-200 citations, and highly cited (201+ citations). 
A significant portion of the studies fall into the lower citations category with 65.21\% of the total studies cited between 1 and 50 times. 
14.49\% of the total studies have a moderate citation count, and 
the 101-200 citations group accounts for only 8.70\% of the total studies. The highly cited group, representing just 4.35\% of the total studies, includes studies that have significantly influenced the field and have been cited more than 200 times. 

\subsection{RQ1 - What are key research works completed to date focusing on Requirements Engineering for software for care of ageing people?}

\subsubsection{RQ1.1 - What is the nature and type of study carried out?}

\begin{table}[h]
\centering
\resizebox{0.8\textwidth}{!}{%
\footnotesize

\begin{tabular}{ll}
\multicolumn{2}{c}{What is the nature/type of the study?}                                                                                                                                                                                                                                                      \\ \hline
\multicolumn{1}{c}{Categories}                                                             & \multicolumn{1}{c}{studies}                                                                                                                                                                                        \\ \hline
\begin{tabular}[c]{@{}l@{}}RE as part of \\ System Development (33)\end{tabular}                & \begin{tabular}[c]{@{}l@{}}S11, S16, S19, S22, S23, S24, S25, S26, S28, \\ S31, S32, S34, S36, S37, S38, S4, S41, S42, \\ S43, S46, S47, S54, S55, S61, S62, S64, S7, \\ S9, S68, S69, S29, S59, S67\end{tabular} \\ \hline
\begin{tabular}[c]{@{}l@{}}Study focusing only on \\ Requirements Gathering  (26)\end{tabular} & \begin{tabular}[c]{@{}l@{}}S1, S12, S14, S17, S18, S2, S20, S23, S3, \\ S30, S35, S40, S44, S48, S49, S5, S50, \\ S52, S6, S60, S63, S8, S9, S39, S21, S66\end{tabular}                                           \\ \hline
\begin{tabular}[c]{@{}l@{}}Study Proposing \\ New/Modified RE Methodology (9)\end{tabular}        & S10, S13, S15, S27, S53, S56, S57, S58, S51                                                                                                                                                                       \\ \hline
Requirements Validation Study (2)                                                              & S24, S45                                                                                                                                                                                                          \\ \hline
RE Guidelines Generation Study (1)                                                             & S65                                                                                                                                                                                                               \\ \hline
RE Exploratory Study (1)                                                                       & S33                                                                                                                                                                                                              
\end{tabular}%
}
\caption{What is the focus/type of the study?}
\label{nature of the study}
\end{table}


We analysed the primary studies in terms of the type of study they present. Table~\ref{nature of the study} presents a summary of the focus of these primary studies. 94.2\% of them focus on RE as a part of overall system development, studies focusing on requirements gathering only, or studies proposing a new or modified RE methodology. Nearly half (33 studies, 47.83\%) of the studies conduct RE as part of System Development i.e. that includes RE tasks with associated system design and implementation tasks. For example, study S22 describes the development of a web-based cognitive training tool for senior people called VIRTRAEL. A user-centred development method was applied to gather and extract older adult users' key needs (requirements). Study S46 presents an application for the development of a concrete system, the “Medication Assistant”, that allows voice and touch interaction to facilitate older adults' access. This study included older end users in their requirements process to identify needs like forgetting to take medication and to increase the usability and accessibility of their product.\\

The second-highest percentage (26 studies, 37.68\%) are studies that focus only on requirements gathering i.e. studies focusing on RE as the key focus without reporting on using the requirements for system development. For instance, study S1  describes a study investigating factors that influence older adults’ intention to use e-health services via a smartphone. It focuses on the process of RE to gather, document, and analyze the key factors in the UTAUT (Unified Theory of Acceptance and Use of Technology) model for older citizens care app. As a result, their study defines that senior citizens are concerned about emotional, social, quality, and price for money when they use the service.\\

A small (9 studies, 13.04\%) percentage of studies focus on proposing a new or modified methodology. These encompass (i) meta-RE studies (e.g., S13, S15, S27), or (ii) new proposed RE design guidelines (e.g., S53, S58). For example, study S13 is a good example of a meta-RE study, which proposes a new ontology-based RE method for well-being, ageing, and health supported by the Internet of Things (IoT). The new ontology RE method in this study makes heart failure patients feel better and cured better as a result. Study S65 presents guidelines for participatory design projects with persons with dementia, benefiting future studies in this area.

A few of the selected primary studies focus on Requirements Validation (2 studies, 2.9\%), Guidelines Generation (1 study), and an RE Exploratory study (1 study). Study S33 is the only example of an `exploratory' RE study. This study discusses the pilot design of a full-scale study to collaborate with older adults but without actual implementation. It highlights the importance of including older stakeholders in the design without any practical design or prototype.  










\subsubsection{RQ1.2 - What aged healthcare and well-being issues are addressed in each work?}

We analysed the primary studies which targeted health and well-being aspects. Table ~\ref{Health aspect of the study} provides an overview of these aspects that were investigated in each study. We classify these into 3 main types of health aspects -- physical challenges (39 studies, 56.52\%), mental challenges (14 studies, 20.29\%), and ageing in general (21 studies, 30.43\%).

\begin{table}[h]
\centering
\resizebox{\textwidth}{!}{%
\begin{tabular}{llll}
\textbf{Main Streams} &
  Health Aspect Major Challenges &
  Health Aspect Minor Challenges &
  Studies \\ \hline
\multirow{12}{*}{\textbf{Physical challenges (39)}} &
  \multirow{5}{*}{Chronic Disease} &
  Chronic Disease(no disease mentioned) &
  S36, S42 \\
 &               & Upper Limb Rehabilitation (ULR) & S18           \\
 &               & Peripheral Arterial Disease     & S42           \\
 &               & Hypertension (Stage 1)          & S44           \\
 &               & Heart Failure                   & S13, S7       \\
 \cmidrule(r){2-4}
 & Acute Disease & Postfracture Acute Pain         & S38           \\
 \cmidrule(r){2-4}
 &
  \multirow{6}{*}{\begin{tabular}[c]{@{}l@{}}Long-term Care \\ challenges\end{tabular}} &
  Living/home Assistance &
  \begin{tabular}[c]{@{}l@{}}S1, S14, S16, S19, S2, S20, \\ S23, S25, S26, S31, S33, S35, \\ S40, S41, S48, S5, S56, S57, \\ S61, S62, S63, S9, S68, S21\end{tabular} \\
 &               & Medication Intake               & S39, S46, S64 \\
 &               & Fall                            & S69, S67      \\
 &               & Disability                      & S59           \\
 &               & Comorbidity                     & S47           \\
 &               & Hearing-aid                     & S66           \\ \hline
\multirow{4}{*}{\textbf{Mental challenges (14)}} &
  \multirow{4}{*}{\begin{tabular}[c]{@{}l@{}}Mental Health or\\ Emotional Challenges\end{tabular}} &
  Cognitive Declines &
  \begin{tabular}[c]{@{}l@{}}S17, S19, S22, S28, S37, S4, \\ S43, S52, S54, S58, S65\end{tabular} \\
 &               & Depression                      & S51           \\
 &               & Loneliness                      & S41           \\
 &               & Parent-Child Relationship    & S3            \\ \hline
\multirow{3}{*}{\textbf{\begin{tabular}[c]{@{}l@{}}Ageing Challenges \\ (in General) (21)\end{tabular}}} &
  \multirow{3}{*}{Geriatric Challenges} &
  Ageing (in general) &
  \begin{tabular}[c]{@{}l@{}}S1, S10, S11, S12, S15, S17, S8,\\ S24, S27, S30, S34, S45, S49, \\ S50, S53, S55, S6, S60, S64,\end{tabular} \\
  \cmidrule(r){3-4}
 &               & Inactivity and Sedentary        & S32           \\
 &               & Frailty                         & S29           \\ \hline
\end{tabular}%
}
\caption{The key aged healthcare and well-being issues are addressed in each work}
\label{Health aspect of the study}
\end{table}

The \emph{physical challenges} studies focus on addressing key health challenges related to chronic challenges (7 studies), acute disease (1 study), and long-term care challenges (32 studies). For example, study S36 explored using virtual communities and mobile technology to improve healthcare for older adults with chronic diseases. Study S38 aimed to develop a mobile app prototype for older adults with osteoporosis using a user-centred design. 

\emph{Mental challenges} are mainly related to mental health or emotional challenges, including 11 studies on cognitive declines, 1 study on depression, 1 study on loneliness, and 1 study on parent-child relationships. For example, in study S51, the key aim is to utilize participatory design (PD) methods to enable older adults diagnosed with depression to actively participate in the design process of socially assistive robots. In study S41, the key aim is to develop and evaluate a smart home technology, SofiHub, designed to support independent living among older adults with loneliness. S3 aims to gather requirements and develop an older adult caring application (Berbakti) to enhance the parent-child relationship in Indonesia. 

\emph{Ageing Challenges (in general)} refers to challenges related to ageing, not specific diseases. For example, study S1 investigated factors influencing older adults' intention to use e-health services via smartphone. Study S32 evaluated the usability, user experience, and effectiveness of the GOAL mHealth intervention for inactive, sedentary older adults. Study S29 developed and tested an online community care platform for frail older adults, focusing on supporting their independence through care, health, and communication functions.

\subsubsection{RQ1.3 - What are the demographics of the older adults in each study?}

\begin{figure}
    \centering
    \includegraphics[width=0.75\textwidth]{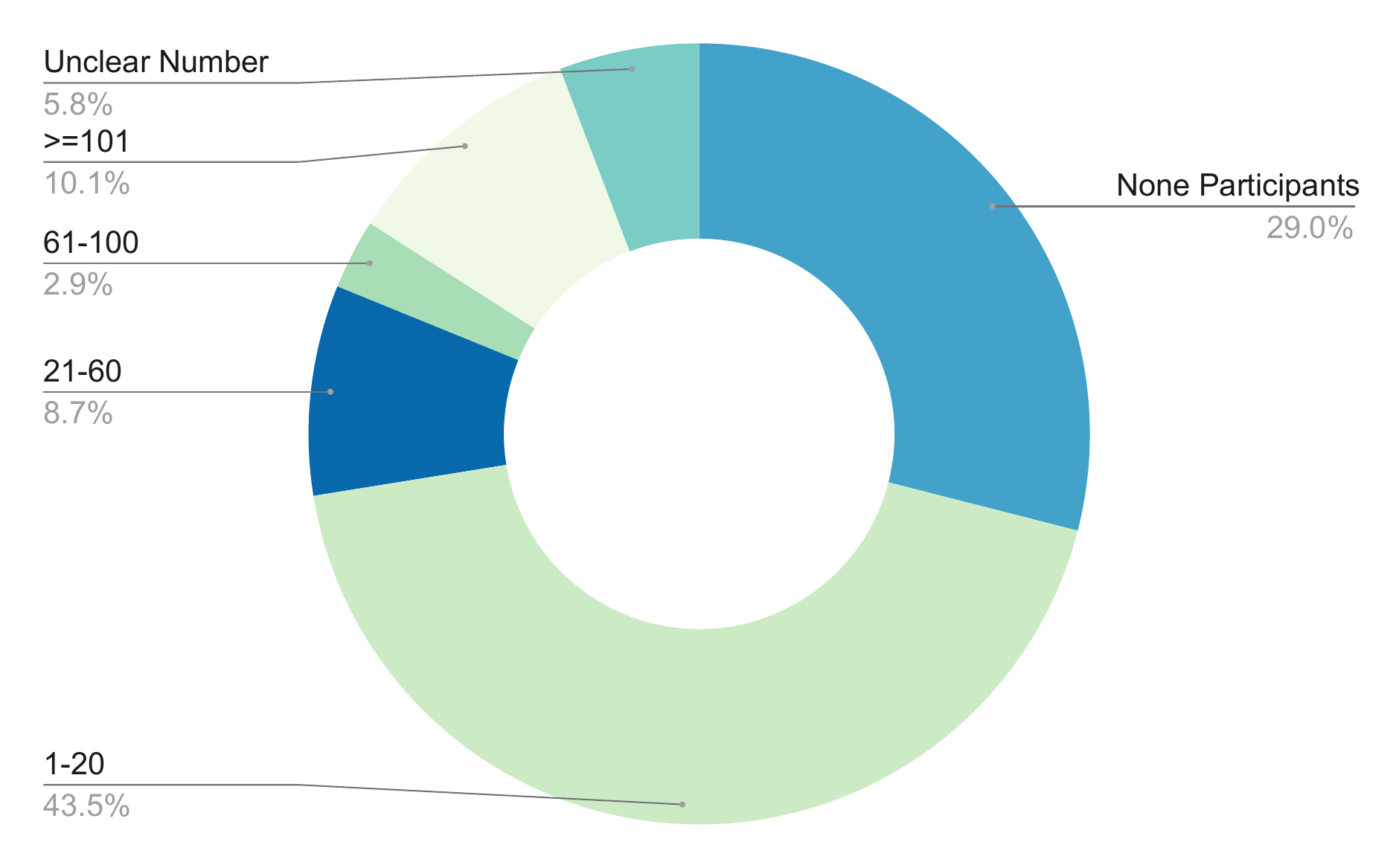}
    \caption{Number of Older Adults Participated in Each Study}
    \label{fig: Num of older adults participated in each study}
\end{figure}


We collected data on the number of older adult participants, their age group distribution, and the average age of participants. If there were any caregivers or technical experts included, we noted their participation as well. We describe each of these findings in detail below. 
\\\\

\noindent \textbf{\textit{Number of Participants: }}
\noindent The number of older adult participants in each study is summarized in Figure ~\ref{fig: Num of older adults participated in each study}. 
We divide number groups into six categories: 1-20 participants, 21-60 participants, 61-100 participants, 101+ participants, unclear (no evidence of participant numbers), and none (clearly stated no participants). The most common category is 1-20 participants, accounting for 43.5\% of the studies. The next largest group is no participants and 101+ participants, both at 10.1\%, followed by 21-60 participants at 8.7\%, and no participants at 5.8\%. The smallest group is 61-100 participants, representing 2.9\%. These variations in sample size can be influenced by factors like the study's target institution size, local population density, funding, etc. Larger participant groups don't necessarily equate to higher participation quality, but they may affect the requirements quality and representation bias. 

\begin{figure}
    \centering
    \includegraphics[width=0.65\textwidth]{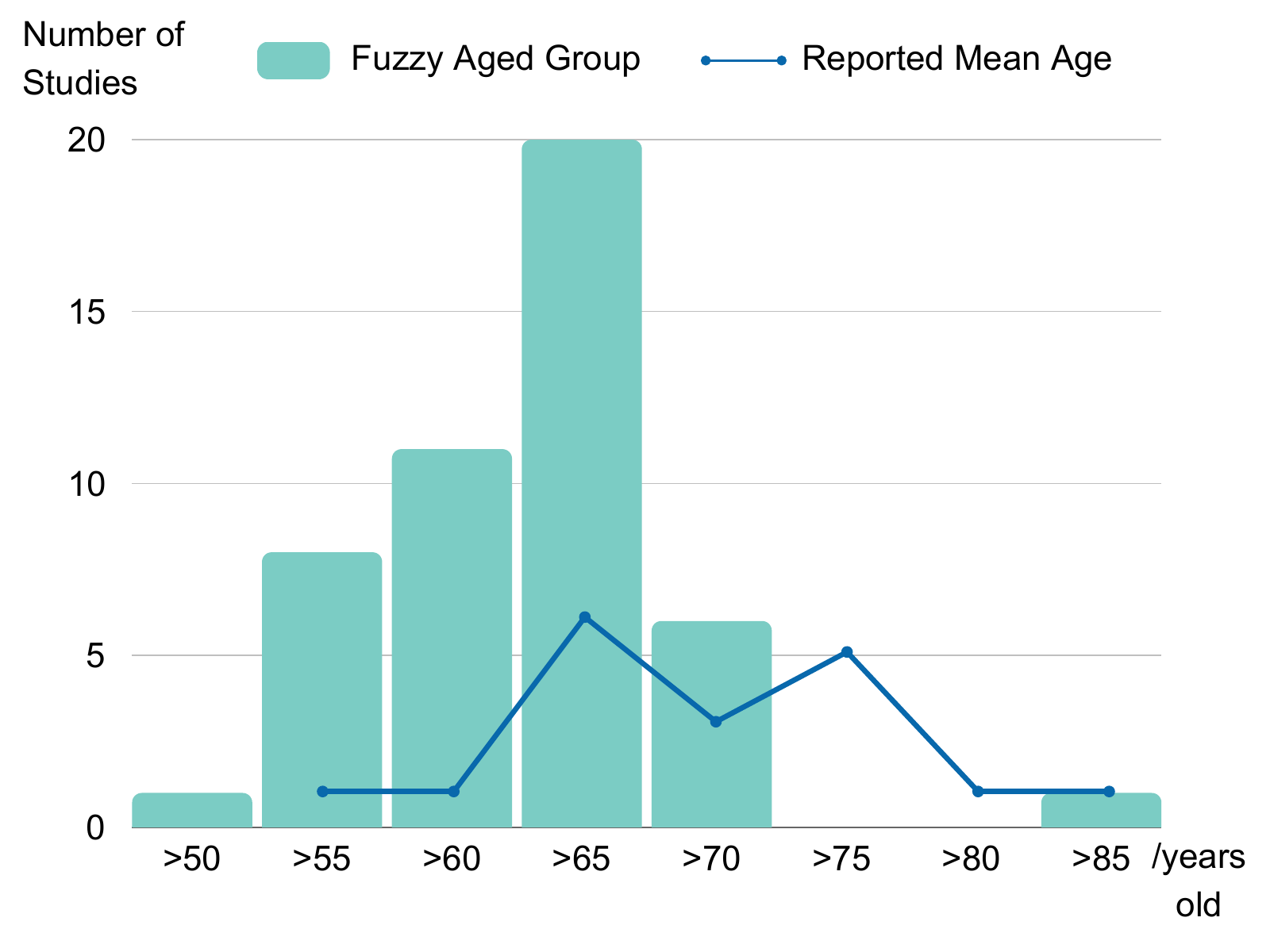}
    \caption{Fuzzy Age Group and Mean Age of Older Adults Participated in Each Study}
    \label{fig: Age of older adults participated in each study}
\end{figure}

\noindent \textbf{\textit{Age Range of Participants:}}
There is no universal definition of `older adults', and the age range can vary by region or country. For instance, individuals aged 65 are often considered older adults in Australia, the UK, and the US, while in China, Thailand, and Malaysia, the threshold is typically 60. In our review, we use the age specified in each study. If no age is mentioned, we consider the commonly accepted older adult age in the country of the study. If participants' ages are unclear and not specified, we leave it unclear.
The age groups and mean ages of older participants in our studies are shown in Figure \ref{fig: Age of older adults participated in each study}. We classified age bands as \(>=\)50 years old, \(>=\)55 years old, \(>=\)60 years old, \(>=\)65 years old, \(>=\)70 years old, \(>=\)75 years old, \(>=\)80 years old, and 85+ years old. We used the reported age range of participants and reported mean age to determine which band it falls into. Most studies involved participants aged \(>=\)65, followed by \(>=\)60, with no studies reporting age groups of \(>=\)75 or \(>=\)80.  However, the mean age data shows that after the 65-70 group, the 75-80 group is the second largest, and the 80-85 group has 1 study. The discrepancy between age groups and mean ages likely reflects the common senior age in many countries, where most studies have participants with a mean age between 65 and 75. The lower numbers in the 80+ group may relate to availability, health concerns, and assistance needs ~\cite{golomb2012older}.

\begin{figure}[h]
   \centering
   \includegraphics[width=0.8\textwidth,trim={0cm 3cm 0cm 3cm},clip]{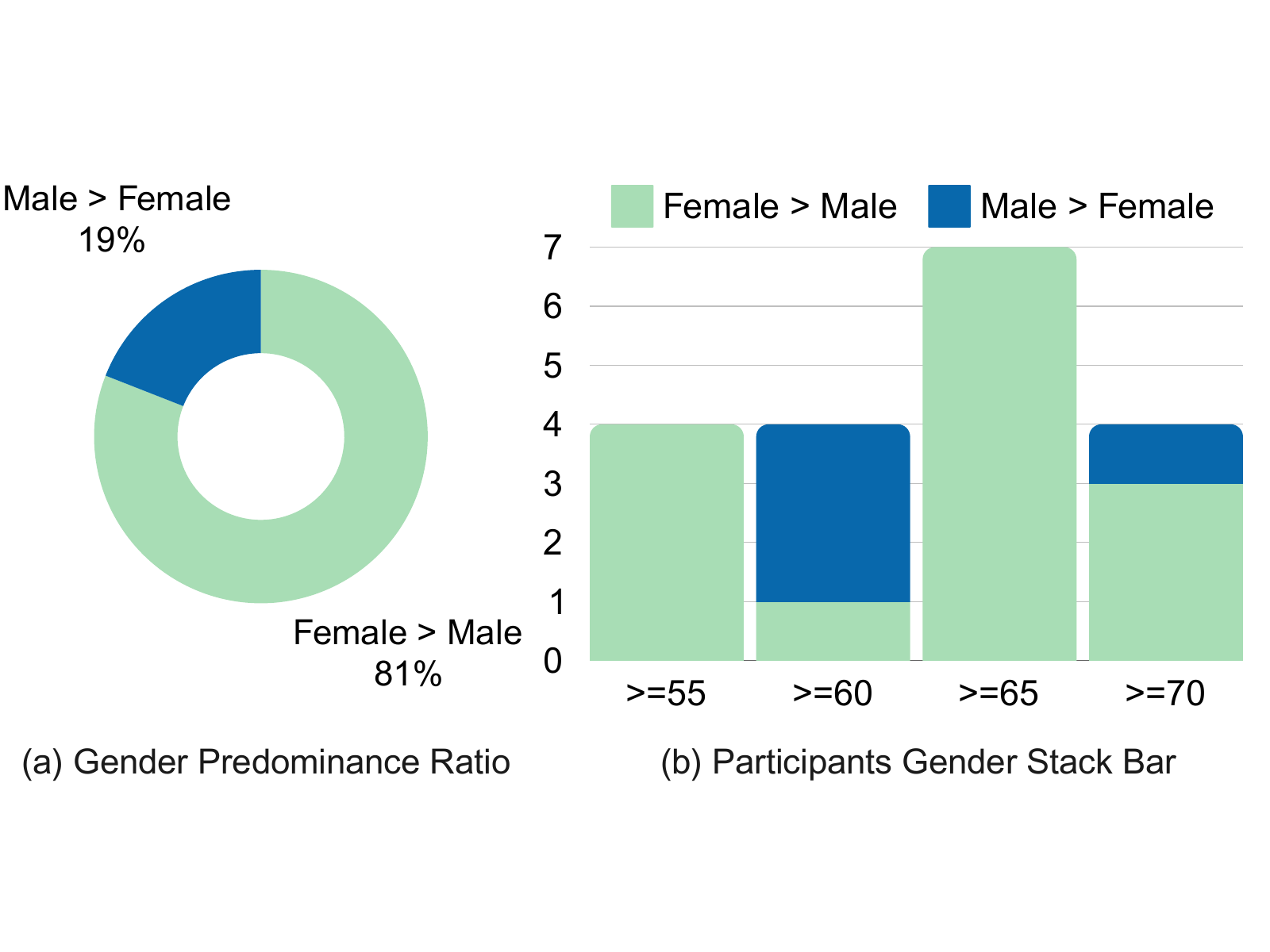}
   \caption{Senior End User Participants Gender Distribution Among Reported Studies}
   \label{fig: senior Participants Gender Distribution}
\end{figure}

\noindent \textbf{\textit{Gender of Participants}}
Figure ~\ref{fig: senior Participants Gender Distribution} summarises the gender distribution of senior user participants in each study if reported. Figure ~\ref{fig: senior Participants Gender Distribution}(a) shows that among all the studies we included, there was a higher representation of more female than male participants, comprising 81\% of the total sample, compared to male participants, who accounted for 19\% of the total sample. Among the studies reporting gender data, there are 17 studies with female-skewed samples, 4 studies with a predominance of male participants, and 48 studies without participant gender information. This gender imbalance with a trend towards greater female participation is aligned with the gender ratio that is commonly observed among the seniors ~\cite{guralnik2000ratio, jensen1994distribution}. 


\begin{figure}[h]
   \centering
   \includegraphics[width=0.8\textwidth]{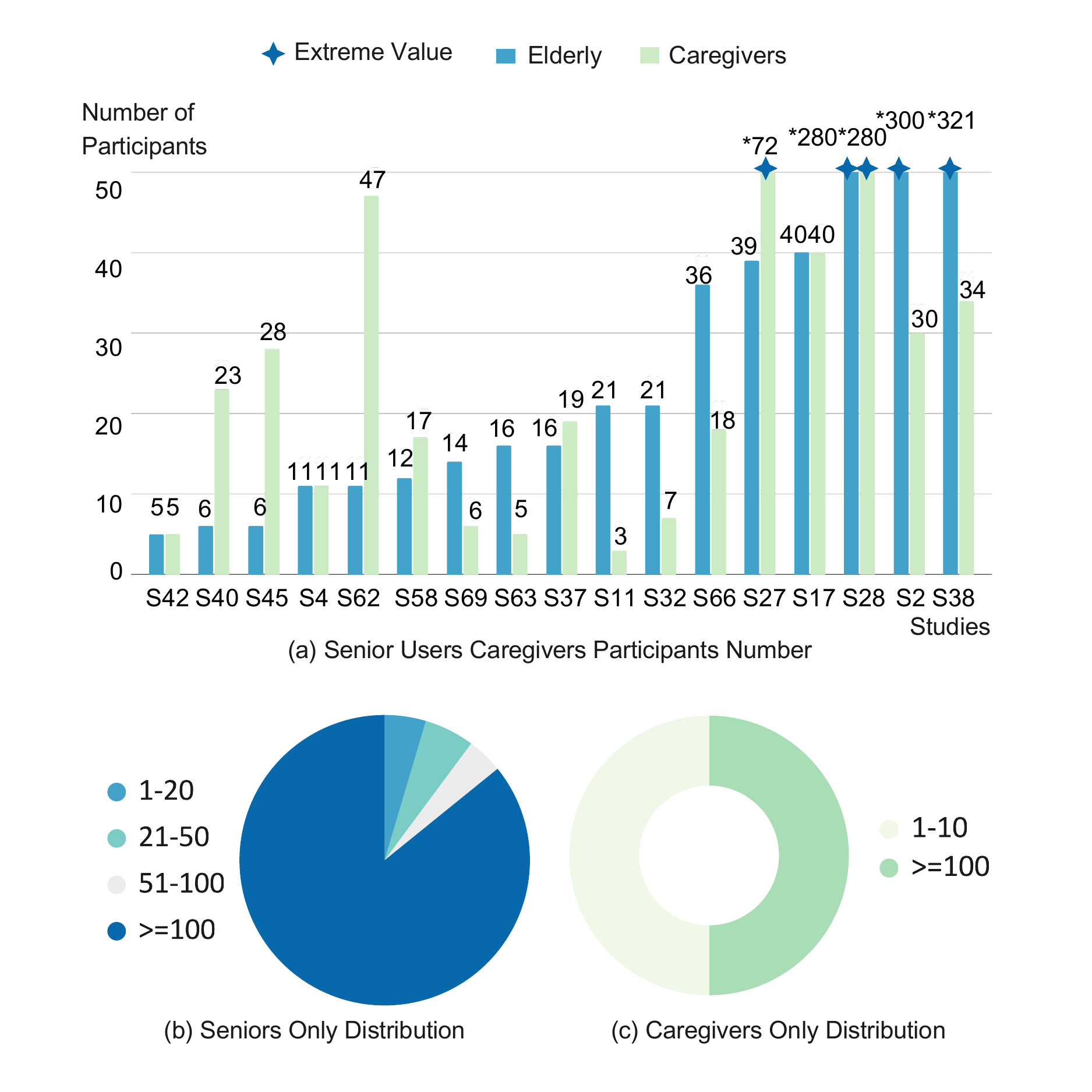}
   \caption{Senior End Users and Caregivers Participated in Each Study}
   \label{fig: Stakeholders participated in each study}
\end{figure}

    
    

\noindent \textbf{\textit{Caregiver Representation: }}
Caregivers in aged care refer to individuals who provide care and support to seniors, including professional caregivers like nurses, clinicians, social workers and informal caregivers like families, volunteers, facility managers, and colleagues. Figure ~\ref{fig: Stakeholders participated in each study}(a) provides the number of study participants that include older adult users and caregivers. The results indicate that in the total number of all included studies, the number of older adult participants is similar to that of caregiver participants. Variations in the ratios of seniors to caregiver participants can be observed across studies with differing participant numbers. Studies with few participants, like S42, S4, and S69, normally included more older adult participants than caregivers. Studies with a medium number of participants, like S62, S45, and S27, normally have more caregivers than older adults. Five studies included more than 100 participants, and the extreme values were labelled with stars in Figure ~\ref{fig: Stakeholders participated in each study}(a). There are no obvious patterns among studies with high total participant numbers like S2, S28, and S38, with more than 280 older adult participants.
In studies S17, S4, and S28, the reason why the caregivers and older adult participants numbers are the same is that they introduced a pair of participants called `dyads', which includes a caregiver and a senior person receiving care to focus on the dynamics of their relationship, the impact of care-giving on both parties and the effectiveness of care-giving strategies.
Figure ~\ref{fig: Stakeholders participated in each study}(b) Figure ~\ref{fig: Stakeholders participated in each study}(c) provides a summary of participants distribution of older adults only and caregivers only in studies. 27 studies included only older adults and 4 studies included only caregivers. 
In the caregivers-only participants chart, half of the studies have more than 100 participants and the other half have only a few participants. We can see that a significant number of studies focusing solely on the older adult population without considering caregivers. 

\begin{figure}
    \centering
    \includegraphics[width=0.7\textwidth]{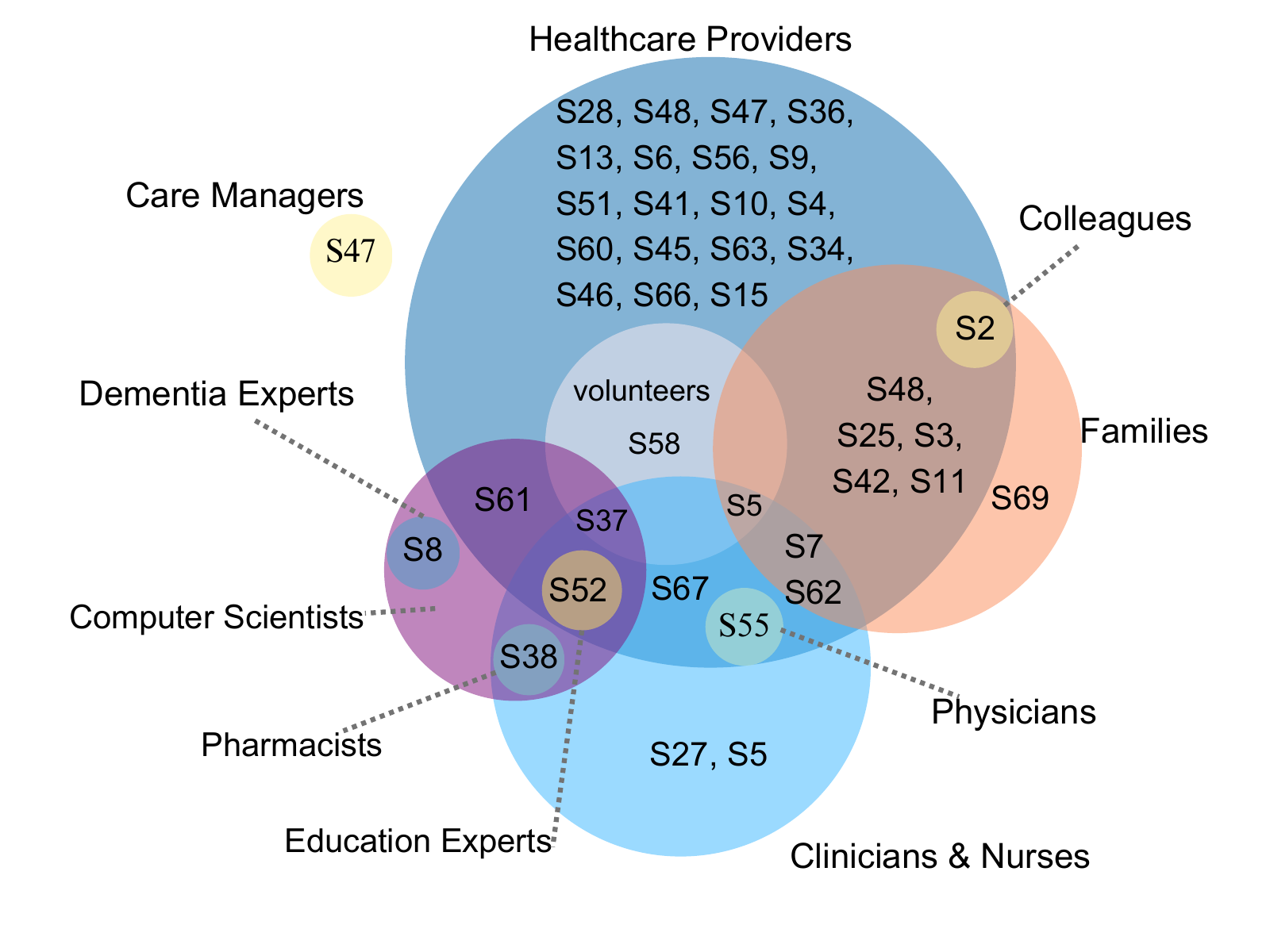}
    \caption{Venn Chart of Participants Types}
    \label{fig: participants types in each study}
\end{figure}

\noindent \textbf{\textit{Participant Types: }} Figure ~\ref{fig: participants types in each study} summarises the different participant types in the primary studies and the studies including single vs multiple types of participants. It can be seen that the most common participant type is general healthcare providers, who are healthcare general practitioners or non-nominate aged caregivers but not clinicians, nurses, and other clinical experts. This suggests the importance of healthcare providers for gathering the needs of older adults in digital health studies. Other significant participant types were families (eg. S25, S3, S42) and clinicians and nurses (eg. S5, S27, S55). 
In some studies (eg. S52, S55, S2), all types of participants are used to understand older adult needs, while in other studies (eg. S47, S69, S28), they only include healthcare providers and older adults as participants. 

\subsubsection{RQ1.4 - What data is captured by the (proposed) software, and how is the data used? What technologies did they use? Does the software use any AI solutions? }
\begin{figure}
    \centering
    \includegraphics[width=0.895\textwidth]{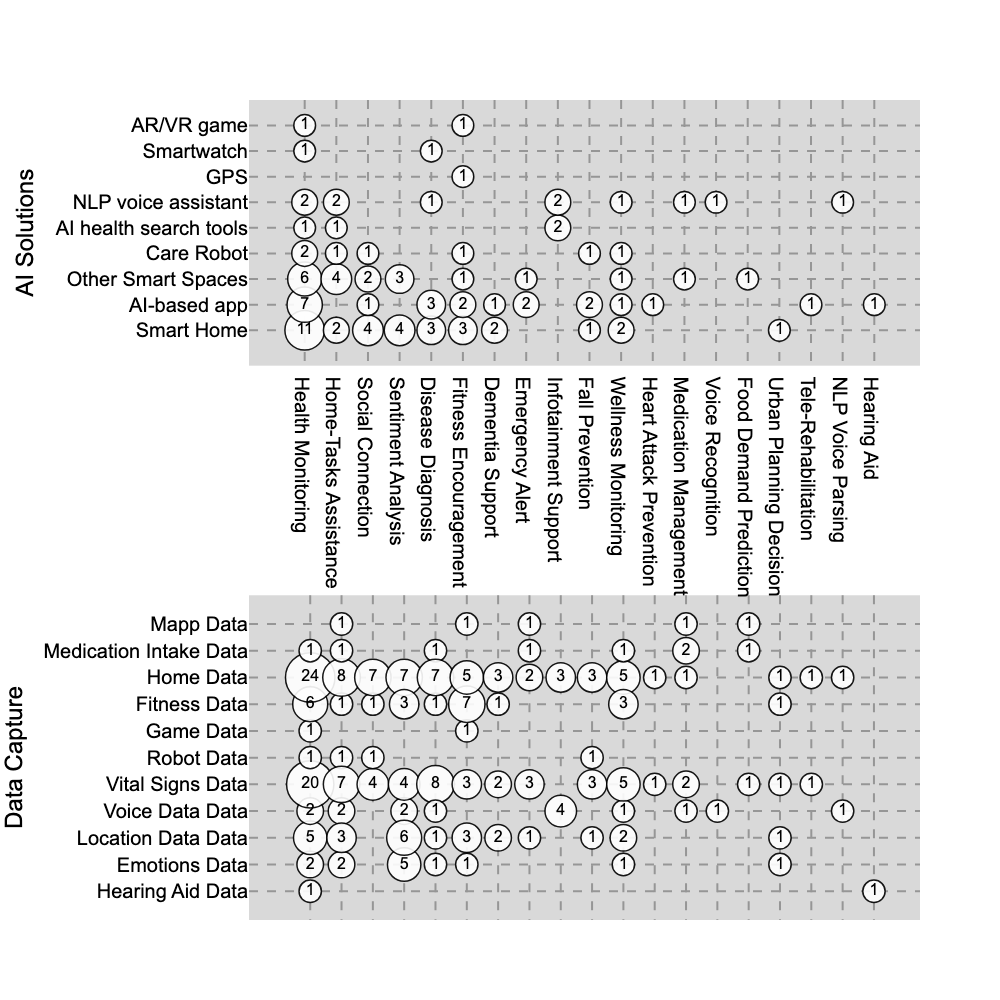}
    \caption{Data Capture Vs AI Usage Vs AI Solutions}
    \label{fig: Data_Capture_Vs_AI_Usage_Vs_AI_Solutions}
\end{figure}

Figure ~\ref{fig: Data_Capture_Vs_AI_Usage_Vs_AI_Solutions} is a double bubble chart that displays the relationship between data capture, AI usage, and AI solutions. The size of the bubble is the number of studies. Data capture refers to the data collection methods and data types in our included studies. AI solutions stand for the applications and technologies that applied AI/ML in each study. AI usage means the functions achieved in each study related to AI/ML algorithms. The x-axis showed the data capture and AI solutions, while the y-axis showed the AI usage that is related to them. Each bubble represents a data point, with the x-axis and y-axis determining its position on the chart and the size of the bubble representing the occurrence of such an x-axis and y-axis combination.

Key AI usage and data capture are home data and vital signs data for health monitoring. Health monitoring in our primary studies is the most common application now for aged care applications with AI/ML. For example, study S25 used MEMSS (micro-electro-mechanical sensor system) to develop a smart home to enhance health monitoring and fitness encouragement. Another important finding is that sentiment analysis, social connection, and wellness monitoring interest many studies. These results may be due to the rising popularity of emotional and wellness care in aged care. For example, study S37 applied AI/ML Algorithms in their dementia support to develop a navigator to support older adults with their dementia challenges and help them improve their quality of life. It is interesting to compare the AI health search tools in AI solutions with the voice NLP questions in data capture which shows it is important and common to apply voice in infotainment support instead of text questions. 


\subsubsection{RQ1.5 - What different human aspects (besides age) were considered in the study (if any)?}
\begin{table}[h]
\centering
\resizebox{0.95\textwidth}{!}{%
\begin{tabular}{lll}
Major Factors &
  \textbf{Minor Factors} &
  \textbf{Papers} \\ \hline
\multirow{5}{*}{\textbf{Personal and Lifestyle (31)}} &
  \begin{tabular}[c]{@{}l@{}}Personal Preferences\\ (eg. food, entertainment, hobbies)\end{tabular} &
  \begin{tabular}[c]{@{}l@{}}S11, S13, S17, S20, S23, S37, S38, S4, S40, S42, S5, \\ S50, S54, S57, S62, S64, S8, S9, S69, S59, S51, S66\end{tabular} \\
 &
  Emotions &
  S1, S12, S13, S17, S38, S4, S41, S5, S62, S64, S7 \\
 &
  Fitness Lifestyle &
  S9, S43 \\
 &
  Personality &
  S10, S64, S8, S44 \\
 &
  Life Experience &
  S10, S6, S53 \\ \hline
\multirow{6}{*}{\textbf{Societal and Environmental (42)}} &
  \begin{tabular}[c]{@{}l@{}}Living/Family Situation \\ or Societal Influences\end{tabular} &
  \begin{tabular}[c]{@{}l@{}}S1, S11, S14, S15, S16, S2, S22, S23, S25, S28, S30, S31, \\ S32, S35, S36, S37, S4, S40, S42, S44, S46, S47, S49, \\ S5, S53, S55, S60, S62, S63, S8, S9, S68, S29, S59, S51\end{tabular} \\
 &
  Culture/Ethnicity &
  S5, S15, S17, S30, S31, S32, S55, S60, S59 \\
 &
  House and Location &
  S11, S2, S53, S56, S64, S8, S9 \\
 &
  Socio-economic Status &
  S1, S12, S20, S3, S30, S31, S32, S53, S8 \\
 &
  Parental Status (Living) &
  S3 \\
 &
  Loneliness &
  S23, S9 \\ \hline
\multirow{8}{*}{\textbf{Health and Well-being (61)}} &
  Physical and Mental Challenges &
  \begin{tabular}[c]{@{}l@{}}S11, S12, S13, S14, S15, S16, S17, S20, S22, S24, S25, S26, \\ S27, S28, S3, S33, S34, S35, S36, S37, S38, S40, S41, S42, \\ S44, S46, S47, S5, S50, S52, S53, S54, S6, S60, S62, S63, \\ S64, S7, S8, S9, S65, S39, S21, S69, S29, S59, S51, S66\end{tabular} \\
 &
  Comfort with Technology &
  \begin{tabular}[c]{@{}l@{}}S1, S10, S11, S12, S13, S2, S20, S22, S26, S28, S32, \\ S34, S35, S36, S38, S40, S42, S43, S45, S46, S47, S49, \\ S53, S54, S56, S63, S64, S8, S39, S29\end{tabular} \\
 &
  Gender &
  \begin{tabular}[c]{@{}l@{}}S20, S22, S28, S30, S31, S32, S35, S40, S42, S44, \\ S50, S53, S55, S60, S62, S63, S29, S59, S66\end{tabular} \\
 &
  Cognitive and Educational Level &
  \begin{tabular}[c]{@{}l@{}}S1, S10, S11, S15, S17, S21, S22, S26, S27, S28, S30, \\S31, S32, S35, S4, S42, S43, S47, S5, S50, S55, S66,\\ S58, S54, S59, S63, S9, S65, S68 \end{tabular} \\
 &
  Occupation and Skill Level &
  S2, S35, S40, S42, S43, S46, S50, S55, S63, S64 \\
 &
  Smoking &
  S42
   \\ \hline
\textbf{Unclear (5)} &
   &
  S19, S18, S48, S61, S67 \\ \hline
\end{tabular}%
}
\caption{Different Human Aspects (Besides Age) Considered In Each Study}
\label{Human aspect of the study}
\end{table}

Table ~\ref{Human aspect of the study} summarises the key human aspects used in RE in our selected primary studies. These are grouped into 3 major categories and 19 minor categories. Out of 69 studies, 5 reported no human factors, 57 used multiple RE techniques, and 7 used only one. Of the studies that reported human factors, 31 studies (44.93\%) addressed personal and lifestyle factors, 42 studies (60.87\%) covered societal and environmental factors, and 61 studies (88.41\%) included health and well-being factors.

\textbf{\emph{Personal and lifestyle factors}} encompass personal preferences (e.g., food, entertainment, clothing, hobbies) used in 22 studies, emotions in 11 studies, fitness lifestyle in 2 studies, personality in 4 studies, and life experience in 3 studies. The most commonly investigated factor is personal preferences. However, it is observed that the collection methods for this factor vary across studies and might be influenced by the provided scenarios. Generally, food, entertainment, and hobbies are the most frequently mentioned factors. For instance, studies S11 and S20 collected data on food preferences and hobbies such as sports, highlighting different aspects of life for end users, which could influence the requirements-gathering process. Emotional and fitness lifestyle factors are often considered in studies involving emotional sensors, incorporating these human factors into the participant group as part of the requirement targets, as demonstrated in studies S62, S64, and S4. 
In study S44, the Bartle Test of Gamer Psychology results indicated the predominant player types in the target group to be explorers and socializers, distinguishing the needs of different elderly individuals in terms of experiencing the game or interacting with friends. Life experience refers to events from the past lives of elderly users, which differ from professional skills or socio-economic status. 

\textbf{\emph{Societal and environmental factors}} include living/family situation or societal influences, used in 35 studies, culture/ethnicity in 9 studies, house and location factors in 7 studies, socio-economic status in 9 studies, parent status in 2 studies, and loneliness in 2 studies. Living/family situations or societal influences refer to any living arrangements, such as living with family or friends or in a residential home with caregivers. 
These factors can significantly influence the needs of older adults, as evidenced in studies S2, S35, and S36. Study S3 revealed that an elderly individual's parents can be crucial to their emotional, social, physical, and cognitive well-being. 

\textbf{\emph{Health and well-being factors}} include physical and mental challenges used by 48 studies, comfort with technology used by 29 studies, gender used by 19 studies, cognitive and educational level used by 28 studies, occupation and skill level used by 10 studies, and smoking used by 1 study. The physical and mental challenges are the most common and can be related to different clinical goals which is easily noticed, especially in studies published on digital health. For example, study S47 found that the clinical goal is an important human factor in evaluating the patient-centred biopsychosocial blended collaborative care pathway for the treatment of multimorbid elderly patients. 
In study S58, patients with different levels of dementia were included and their needs were gathered by their dementia clinicians and caregivers.
Smoking is used by study S42 as a crucial human factor in a health self-management system for Peripheral Arterial Disease patients and it is shown that smoking can influence their clinical situation and hence influence the needs of patients in the system.

\begin{figure}[!ht]
    \centering
    \includegraphics[width=0.785\textwidth]{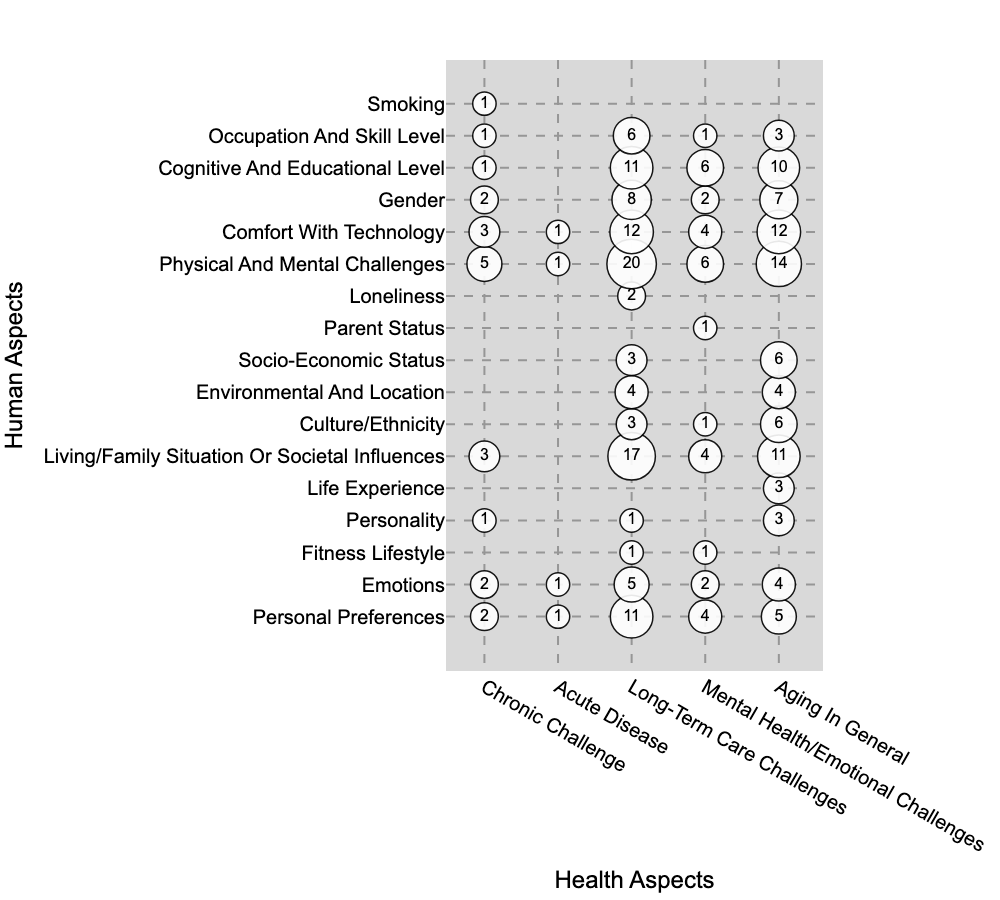}
    \caption{Human Aspects Vs Health Aspects}
    \label{fig:humanVshealth}
\end{figure}

In Figure ~\ref{fig:humanVshealth}, we analysed health aspects and corresponding human aspects of each study in the research. The size of each bubble correlates with the number of studies. This reveals a rough trend where more studies are generally associated with popular human aspects (eg. physical and mental challenges, living situations, and personal preference) in all health aspects. the large bubble size for the cognitive and educational levels paired with mental health or emotional challenges highlights the importance of the human factor ``cognitive level" in emotional challenge health aspect. For the ``ageing in general" category, it is obvious that ``societal and environmental" and ``health and well-being" are investigated more than ``Personal and lifestyle" human factors. This inconsistency may mean that researchers have not yet explored the personal and lifestyle human factors enough for general ageing digital health systems.

\begin{center}
\begin{myframe}[\centering\textbf{RQ1 Answers}]
\footnotesize
Half of our selected primary studies are on the development of a digital health system for older adults with a detailed RE process, about one-third propose a method of RE for such systems, and the rest are guidelines for RE in such scenarios. Half of the studies focus on ageing in general or the comorbidity of older adults in the process of ageing. The rest focus on supporting mental health and specific physical challenges of ageing people. The majority of studies included older adults as participants, but the age group of older adults can vary based on the research topic and the country in which the study was conducted. More than half of the studies included caregivers as participants in their study, and a few studies included developers or technical experts as participants. One-third of studies applied AI/ML in their systems to build the system and collect data, especially after 2015. Diverse human aspects have been considered, including common ones like living situations, and cognitive status, and uncommon ones like loneliness and personality.
\end{myframe}
\end{center}

\subsection{RQ2 - How was the Requirements Engineering carried out?}
\subsubsection{RQ2.1 - What RE techniques did each study use?}

Table ~\ref{RE techniques of the study} summarises the key RE techniques used in our selected primary studies. These are grouped into nine major categories and fourteen minor categories. Of the 69 studies included, 8 reported no RE techniques, 44 used multiple techniques, and 17 focused on a single technique. 

\begin{table}[h]
\centering
\resizebox{\textwidth}{!}{%
\begin{tabular}{@{}lll@{}}
\toprule
\textbf{Major Categories} & \textbf{Minor Categories} & \textbf{Studies}                                                            \\ \midrule
\textbf{Interviews (36)} && \begin{tabular}[c]{@{}l@{}}S10, S11, S13, S14, S17, S19, S23, S27, S28, S31, S32, S36, \\ S37, S41, S42, S44, S46, S5, S50, S53, S55, S57, S6, S60, \\ S61, S62, S63, S64, S7, S39, S68, S21, S69, S29, S59, S51\end{tabular} \\ \midrule
\textbf{Surveys (26)} & & \begin{tabular}[c]{@{}l@{}}S11, S12, S2, S20, S22, S24, S27, S28, S3, S30, S31, S32,\\ S34, S35, S38, S41, S44, S50, S52, S53, S57, S60, S64, \\ S8, S39, S69\end{tabular}\\ \midrule
\textbf{Workshops (23)} && \begin{tabular}[c]{@{}l@{}}S10, S14, S20, S27, S28, S35, S36, S37, S38, S4, S40, \\ S49, S50, S53, S57, S58, S61, S9, S39, S69, S59, S51, S66\end{tabular}  \\ \midrule
\textbf{Observations (21)} & & \begin{tabular}[c]{@{}l@{}}S10, S11, S13, S17, S19, S22, S28, S46, S5, S54, \\ S57, S58, S61, S62, S64, S7, S68, S69, S29, S51, S66\end{tabular} \\ \midrule
\multirow{4}{*}{\textbf{Material analysis (11)}}  & Document analysis   & S32, S36, S40, S44, S53, S58, S61, S63, S7, S65   \\
& Video analysis & S28 \\
& audio analysis& S7, S32\\
& \begin{tabular}[c]{@{}l@{}}scenario-based user need analysis\\  (SUNA)\end{tabular} & S36\\ \midrule
\textbf{Rapid Prototyping}&  & S10, S17, S2, S22, S63, S64, S68, S21, S69  \\ \midrule
\multirow{3}{*}{\textbf{Innovation process (6)}} & Brainstorming& S33, S46, S49, S52, S64\\
& Design Thinking Process & S47 \\
 & Enlisting Allies in Recruitment  & S49  \\ \midrule
\textbf{Psychology Test (1)} & Bartle Test of Gamer Psychology  & S44  \\ \midrule
\textbf{Reuse existing requirements} &   & S1, S17, S34, S43, S45, S52 \\ \midrule
\textbf{Unclear}  &  & S15, S16, S18, S25, S26, S48, S56, S67  \\ \bottomrule
\end{tabular}%
}
\caption{The RE techniques used in each study}
\label{RE techniques of the study}
\end{table}

Of the studies that reported their RE techniques (including overlaps), 36 studies (52.17\%) used interviews, 26 studies (37.68\%) used surveys, 23 studies (33.33\%) used workshops, 21 studies (30.43\%) used observations, 11 studies (15.94\%) used material analysis, 9 studies (13.04\%) used rapid prototyping, 6 studies (8.7\%) used innovation process, 1 study (1.45\%) used a psychological test and 6 studies (8.7\%) reused existing requirements. 
Material analysis included document analysis (10 studies), video analysis (1 study), audio analysis (2 studies), and scenario-based user need analysis (SUNA) (1 study). 
Innovation processes included brainstorming (5 studies), design thinking (1 study), and enlisting allies in recruitment (1 study). 
Some studies reused existing requirements. For example, S43 improved a memory game using age-related requirements from previous works by the same authors ~\cite{kawamoto2013application,kawamoto2014converging}.


\subsubsection{RQ2.2 - What tools were used for requirements elicitation and documentation in each study?}

\begin{table}[h]
\centering
\resizebox{\textwidth}{!}{%
\begin{tabular}{lll}
\textbf{Major Categories} & \textbf{Minor Categories} & \textbf{Studies} \\ \hline
\multirow{2}{*}{\textbf{RE tools for documentation (32)}} &
  Text(Word/Docs)/Table(Excel/Sheets) &
  \begin{tabular}[c]{@{}l@{}}S1, S31, S32, S36, S37, S38, S40, S42, S44, \\ S47, S5, S50, S53, S55, S56, S58, S60, S61, \\ S64, S7, S39, S59, S66\end{tabular} \\
 &
  \begin{tabular}[c]{@{}l@{}}Prototyping Tools\\ (eg. Axure RP,or Sketch)\end{tabular} &
  \begin{tabular}[c]{@{}l@{}}S2, S20, S23, S24, S25, S3, S32, S38, S42, \\ S46, S47, S5, S53, S55, S58, S61, S64, \\ S39, S68, S29, S59, S66\end{tabular} \\ \hline
\multirow{5}{*}{\textbf{RE tools for management (17)}} 
 &
  \begin{tabular}[c]{@{}l@{}}Graphical Specification Language Models \\ (eg.UML model tools, meta models)\end{tabular} &
  \begin{tabular}[c]{@{}l@{}}S17, S18, S2, S36, S38, S41, S48, S55, S56, \\ S13, S62, S8, S29, S67, S66\end{tabular} \\
 &
  socio-technical method(eg. Work system design WSD) &
  S7, S47, S48, S67 \\
 &
  Application Lifecycle Management (ALM) &
  S64, S29 \\ \hline
\multirow{4}{*}{\textbf{RE tools for analyzing (5)}} &
  Issue and Project Tracking Tools(eg. Jira,Trello,or Asana) &
  S30 \\
 &
  Aspect-Oriented Requirement Engineering (AORE) &
  S25 \\
 &
  Judgement Call &
  S14 \\
 &
  Data analysis software (eg. NVivo, ATLAS.ti 8) &
  S44, S62 \\ \hline
\textbf{Unclear (26)} &
   &
  \begin{tabular}[c]{@{}l@{}}S10, S11, S12, S15, S16, S19, S22, S26, S27, \\ S28, S33, S34, S35, S4, S43, S45, S49, \\ S52, S54, S57, S63, S9, S65, S21, S69, S51\end{tabular} \\ \hline
\end{tabular}%
}
\caption{The tools used during requirements elicitation and modelling in each study}
\label{RE Tools of the study}
\end{table}

Table ~\ref{RE Tools of the study} presents three categories of Requirements Engineering (RE) tools used in the primary studies. These we grouped into requirements documentation (32 studies, 46.38\%), management (17 studies, 24.64\%), and analysis (5 studies, 7.25\%). The category ``unclear"  includes 26 studies without specific RE tools mentioned. We believe some studies generated user stories likely used RE tools like interviews as well, though they did not specify which ones. Of the 43 studies that reported RE tools, 35 used multiple tools, while 8 used only one.

The dominant category we found was RE tools for documentation. This includes two major subcategories: \emph{Text} and \emph{Tables} tools (23 studies using), and \emph{Prototyping} tools (13 studies using). 
The second largest category we found was RE management tools, including Abstraction Modelling tools, Application Lifecycle Management (ALM), etc. \emph{Graphical specification language} models (eg. UML model tools, meta models) are used in 15 studies. 
\emph{Socio-technical methods} is used in 4 studies. For example, in study S48, Work system design is used to represent the needs of older adults and care managers in an evaluation of a patient-centred biopsychosocial blended collaborative care pathway for the treatment of multimorbid elderly patients (ESCAPE). 2 studies used \emph{Application Lifecycle Management} (ALM) tools. 




The RE tools for analysis category includes 4 sub-categories of tools, each linked to specific studies where they were utilized. \emph{Issue and Project Tracking Tools} are used by a single study, study S30. Confirmatory Factor Analysis (CFA) and Structural Equation Modelling (SEM) are used to analyse the requirements of older adults to understand the end user perspective of smart homes for elderly healthcare. \emph{Aspect-oriented requirements engineering} (AORE) is used by one study, S25. 
\emph{Judgment Call} was created for industry product teams so that ethical concerns could surface when developing Artificial Intelligence (AI) technology, used in study S14. 
\emph{Data analysis} software was used in 2 studies. 

\subsubsection{RQ2.3 -  How were the requirements modelled in each study? }
\begin{table}[ht]
\centering
\resizebox{\textwidth}{!}{%
\begin{tabular}{lll}
\textbf{Major Categories} & \textbf{Minor Categories} & \textbf{Studies}                            \\ \hline
\multirow{4}{*}{\textbf{user centred Design UCD (59)}} 
&
  User stories &
  \begin{tabular}[c]{@{}l@{}}S12, S13, S14, S16, S19, S2, S22, S24, S25, S27, S28, S3, \\S32, S34, S36, S37, S38, S4, S41, S43, S44, S46, S48, S49, \\S5, S53, S54, S58, S61, S62, S63, S64, S7, S9, S68, S21, \\S69, S29, S59, S51, S66\end{tabular} \\
 &Use Case Modelling &
  \begin{tabular}[c]{@{}l@{}}S1, S10, S11, S13, S16, S14, S17, S19, S2, S20, S25, S29, \\S3, S30, S31, S37, S38, S40, S4, S41, S43, S44, S46, S47, \\S48 ,S5, S50, S53, S55, S58, S63, S64, S68, S66, S67, S56, \\S7, S8, S9\end{tabular} \\ 
 & Personas &
  \begin{tabular}[c]{@{}l@{}}S15, S26, S35, S42, S46, S48, S5, S55, S62, S64, S7, S9, \\ S68, S59\end{tabular} \\
& Goal Model(e.g. i-Star,KAOS) & S11, S2, S41, S62, S63, S7, S8    \\
& fictional user               & S47                               \\ \hline
\multirow{8}{*}{\textbf{Semantic Model (24)}} &
  Activity Diagram &
  S15, S17, S2, S26, S41, S42, S58, S61, S9, S67 \\
                 & Affinity Diagram          & S14, S17, S49, S50, S54, S9       \\
                 & Conceptual Model & S11, S16, S3, S8, S6, S27       \\
                 & Meta model                   & S16, S18, S41, S53                \\
                 & Ontology modelling           & S32                               \\
                 &Comprehensive model &S64 \\ \hline
\textbf{Unclear (7)}& & S23, S33, S45, S52, S57, S60, S39 \\ \hline
\end{tabular}%
}
\caption{How the requirements are modelled}
\label{Requirements models of the study}
\end{table}

Table ~\ref{Requirements models of the study} provides a categorisation of the main types of Requirements Engineering (RE) models used, including 59 studies (85.51\%) that used user-centred design (UCD), 24 studies (34.78\%) that used semantic model, and 7 studies (10.14\%) did not report a specific RE model they used. Of the 62 reported studies, 44 used multiple RE models, and 18 used a single model.

Of the studies that used some form of user-centred modelling, 41 studies used \emph{user stories}, 39 studies used \emph{use cases} to model requirements, 14 studies used \emph{personas}, 7 studies used \emph{goal modelling}, and 1 study used a \emph{fictional user}. User stories are short, simple descriptions of a feature or functionality from the perspective of an end user. 
Personas are fictional characters created to represent different user types or user roles that might interact with a system based on real user data. Goal modelling is a technique used in RE to capture and represent the goals and objectives of stakeholders. 
A fictional user is a character created for the purpose of testing or illustrating a system, but different from personas, they are not based on real data or research but are used to simulate real user interactions and scenarios. 

Semantic models were classified into 6 minor categories. \emph{Activity diagrams} are used by 10 studies, which represent the flow of control or the sequence of activities in a system or process. \emph{Affinity diagrams} are used by 6 studies, which organize a large number of ideas, issues, or information into groups based on their natural relationships. 
\emph{Conceptual models} are described in 6 studies. These provide a high-level representation of a system to communicate the overall design or structure of a system without implementation details. 
A \emph{meta-model} defines the structure and semantics of other models used to represent their requirements, which is used by 4 studies. 
An \emph{ontology model} was used by one study.
Finally, a \emph{comprehensive model}, for example, an International Classification of Functioning Disability and Health (ICF) model, was used by one study, S64. 

\subsubsection{RQ2.4 - How were the requirements validated?}
\begin{table}[h]
\centering
\resizebox{\textwidth}{!}{%
\begin{tabular}{lll}
\hline
\textbf{Validation Major Categories} &
\textbf{Validation Minor Categories} &
\textbf{Studies} \\ \hline
\textbf{\begin{tabular}[c]{@{}l@{}}Prototyping \\ \&\\ Proof of Concept (30) \end{tabular}} &    &   \begin{tabular}[c]{@{}l@{}}S10, S11, S13, S19, S20, S23, S24, S32, S34, S37, \\ S38, S42, S47, S49, S5, S53, S54, S55, S57, S58, \\ S6, S62, S64, S9, S68, S69, S29, S59, S51, S66\end{tabular} \\ \hline
\multirow{4}{*}{\textbf{Reviews by Human (37)}} &
  \begin{tabular}[c]{@{}l@{}}Stakeholder feedback\\ (eg.workshops)\end{tabular} &
  \begin{tabular}[c]{@{}l@{}}S11, S12, S17, S19, S22, S27, S32, S36, S37, S38, \\ S40, S41, S43, S44, S45, S46, S49, S5, S54, S58, \\ S6, S60, S64, S21, S69, S29, S51, S66\end{tabular} \\
 &
  Requirements Reviews/Inspections &
  \begin{tabular}[c]{@{}l@{}}S10, S13, S24, S31, S35, S5, S54, S55, S64, S7, \\ S9, S21, S51\end{tabular} \\
 &
  External Reviews &
  S28, S35, S46, S5, S44 \\ \hline
\multirow{2}{*}{\textbf{Testing by Human (16)}} &
  Scenario Testing &
  \begin{tabular}[c]{@{}l@{}}S20, S28, S36, S38, S42, S47, S48, S61, S62, \\ S64, S68, S59\end{tabular} \\
 &
  Use Cases Validation (eg.input/output) &
  S16, S26, S48, S7, S67 \\ \hline
\textbf{Automated Testing (11)} &
  User Acceptance Testing (UAT) &
  S10, S22, S27, S28, S30, S31, S32, S34, S38, S54, S64 \\\hline
\textbf{Score/Scale system (2)} &
   &S8, S32 \\\hline
\textbf{Unclear (15)} &
   &
  \begin{tabular}[c]{@{}l@{}}S1, S14, S15, S18, S2, S25, S3, S33, S4, S50, \\ S52, S56, S63, S65, S39\end{tabular} \\ \hline
\end{tabular}%
}
\caption{How the requirements were validated in the study}
\label{RE Validations of the study}
\end{table}

Table ~\ref{RE Validations of the study} categorises our primary studies into five main validation methods: prototyping (30 studies, 43.48\%), reviewing by humans (37 studies, 53.62\%), testing by humans (16 studies, 23.19\%), automated testing (11 studies, 15.94\%) and scoring(2 studies, 2.9\%) in the context of software engineering research. The `unclear/none' category (15 studies, 21.74\%) includes studies where the specific RE validation tools were unclear or no validation process was mentioned.
Of the 54 studies reporting validation methods, 38 used multiple methods, while 16 used only one. 

Prototyping and Proof of Concept involves creating experimental software versions to showcase key features and gather stakeholder feedback. For example, study S10 developed a prototype to understand senior users' needs for the SeniorDT RE framework. The largest category, Reviews/Validation by Humans, includes the process of validation by humans checking that the software meets the specified requirements and satisfies the needs of stakeholders, including \emph{stakeholder feedback} (28 studies), \emph{requirements reviews/inspections} (13 studies), and \emph{external reviews} (5 studies).
External reviews refer to evaluations or assessments of a product, process, or system conducted by individuals or organizations outside of the entity responsible for the item being reviewed. 

Testing by humans involves manually executing test cases to verify the correctness and quality of a software system. Two main approaches described in studies were \emph{scenario testing} (13 studies) and \emph{use case validation} (5 studies). 
The automated testing category includes studies that conducted automated tests. User acceptance testing (UAT) was used in 11 studies. 
For example, Study S10 used requirements testing to enhance older participants' engagement and gather insights using the SeniorDT framework.



\subsubsection{RQ2.5 - Were the requirements used to build an actual system? If so, in which SE stages?}
\begin{table}[h]
\centering
\resizebox{\textwidth}{!}{%
\begin{tabular}{@{}lll@{}}
\toprule
\textbf{Major Stage}& \textbf{Minor Stage}& \textbf{Studies}                                                                                                                                                                                                                                                                                                                 \\ \midrule
\multirow{4}{*}{\textbf{Preparation (62)}}     
 & \begin{tabular}[c]{@{}l@{}}Elicitation\end{tabular} & \begin{tabular}[c]{@{}l@{}}
                                 S11, S13, S14, S15, S16, S17, S19, S18, S2, S20, S21, S22, S23, S24, S25,\\ 
                                 S27, S3, S31, S32, S34, S36, S37, S38, S4, S41, S42, S43, S44, S45, S46,\\
                                 S47, S48, S49, S5, S50, S52, S53, S54, S55, S58, S39, S69, S59, S6, S60,\\
                                 S62, S7, S8, S9, S51, S66\end{tabular}                    \\
& Documentation &                S2, S32, S35, S45, S49, S5, S55, S58, S6, S60, S63, S64, S8, S21, S66 \\
& Analysis  & \begin{tabular}[c]{@{}l@{}}
                                 S15, S2, S26, S27, S32, S35, S36, S42, S48, S49, S5, S50, S52, S57, S58, \\
                                 S60, S61, S63, S64, S8, S68, S67, S51, S66\end{tabular} 
                                 \\
& Design  & \begin{tabular}[c]{@{}l@{}}
                                 S11, S13, S15, S16, S17, S18, S2, S20, S22, S23, S24, S25, S26, S27, S28, \\
                                 S3, S33, S34, S35, S36, S38, S4, S40, S41, S42, S43, S44, S45, S46, S47, \\
                                 S48, S5, S50, S53, S54, S55, S57, S58, S6, S60, S61, S62, S63, S64, S7, \\
                                 S9, S39, S68, S21, S69, S59, S67, S51, S66
                                 \end{tabular}                                                                                                                                                         \\ \midrule
\textbf{Development (37)}                      &                                                                   & \begin{tabular}[c]{@{}l@{}}
S13, S16, S22, S25, S26, S27, S31, S32, S33, S35, S36, S37, S38, S4, S40,\\
S42, S43, S44, S47, S5, S50, S53, S54, S58, S6, S60, S61, S62, S63, S64, \\
S7, S9, S68, S69, S59, S67, S51\end{tabular} \\ \midrule
\multirow{2}{*}{\textbf{Verification (41)}}    & Validation                                                                 & \begin{tabular}[c]{@{}l@{}}
S12, S13, S16, S17, S2, S20, S24, S26, S27, S28, S30, S32, S34, S36, S37, \\
S38, S40, S41, S42, S43, S45, S46, S47, S49, S5, S50, S58, S6, S62, S64, \\
S7, S9, S21, S69, S29, S59, S67, S66\end{tabular}                                                                                  \\ & Testing & S1, S19, S20, S22, S24, S28, S30, S38, S5, S54, S64, S29, S59, S67 \\ \midrule
\multirow{2}{*}{\textbf{Maintenance (7)}} & Updates& S11, S2, S53, S64, S39, S51\\
& \begin{tabular}[c]{@{}l@{}}Deployment\end{tabular}         & S2, S39, S29\\ \midrule
\textbf{Unclear (3)}                          &                                                                              & S10, S56, S65 \\ \bottomrule
\end{tabular}%
}
\caption{In which parts of the process of system building the requirements were used}
\label{RE Used Stage of the study}
\end{table}

Table ~\ref{RE Used Stage of the study} highlights key SE stages where the requirements were used including preparation (62 studies, 89.86\%), development (37 studies, 53.62\%), verification (41 studies, 59.42\%), and post-production (7 studies, 10.14\%). Of 66 studies reporting SE stages, 64 involved multiple stages, while 2 applied them to only one stage. Three studies did not specify any SE stage or activities. 

Preparation activities include \emph{elicitation} (51 studies), \emph{documentation} (15 studies), \emph{analysis} (24 studies), and \emph{design} (54 studies). Elicitation involves gathering and prioritizing requirements from stakeholders. For example, S14 elicits requirements from ethical considerations when developing AI technology for older adults in long-term care settings, and S52 prioritizes the needs of older adults using situation-aware mobile devices. Documentation entails creating documents that describe software requirements and design; in S58, notes and audio recordings help extract requirements and update the iterative prototype. 
Analysis activities include analyzing requirements and system architecture. S33 visualizes biomechanical data on the functional demand of older adults, and S57 analyses the observations and survey data to extract the requirements of stakeholders. Design activities involve conceptualizing software architecture and user interfaces, 

The development stage involves implementing the software based on requirements. For example, S53 built a prototype fostering empathy and individualized design to reduce abstract thought, while S68 developed the telecare app CareMe to improve health monitoring and social connection for older adults. 

The Verification stage includes 38 studies that focused on \emph{requirements validation} and 15 studies that focused on \emph{requirements-based testing}. Validation involves ensuring that the software meets the specified requirements and testing activities to ensure that the software functions correctly. For example, S29 developed an online community care platform for frail older adults, using observations and interviews to assess and refine user requirements for platform modifications. 

Maintenance includes \emph{software updates} (6 studies) and \emph{deployment} (3 studies). Updates may be made to the software based on user feedback or new requirements. 
For example, S64 emphasized user testing and feedback from older adults during development, highlighting how feedback improved their medication assistant app. Deployment \& Maintenance involve launching the software and ensuring its continued functionality over time. For example, S39 demonstrated how requirements were applied during the maintenance of a home medication adherence assistant. 

\subsubsection{RQ2.6 - Were the requirements used to evaluate the solution? If so, how the study is evaluated?}

\begin{table}[h]
\centering
\resizebox{\textwidth}{!}{%
\begin{tabular}{@{}lll@{}}
\textbf{Evaluation Method}& & \textbf{Studies}  \\ \midrule
\multirow{3}{*}{\textbf{\begin{tabular}[c]{@{}l@{}}End User \\ Evaluation (50)\end{tabular}}}      & \multirow{2}{*}{\begin{tabular}[c]{@{}l@{}}Real-world \\ End User Testing\end{tabular}}   
&  \begin{tabular}[c]{@{}l@{}}
S6, S9, S10, S11, S12, S13, S17, S19, S20, S21, S22, \\ S24, S27, S28, S29, S3, S30, S32, S34, S35, S36, \\ S38, S4, S40, S41, S43, S45, S46, S47, S5, S53, S54, \\S58, S59, S60, S62, S64, S65, S66, S68, S69\end{tabular} \\ 
& Existing User Data Evaluation & S39, S61 \\ \
& \begin{tabular}[c]{@{}l@{}}User Requirement \\ Reviews\end{tabular} & S7, S48, S9, S11, S16, S26, S38, S42, S62, S69, S67  \\ \midrule
\multirow{2}{*}{\textbf{\begin{tabular}[c]{@{}l@{}}Domain-Specific \\  Evaluation (7)\end{tabular}}} 
& Domain Testing    & S32, S29        \\   

& Domain Analysis   & S46, S24, S53, S51, S44  \\ \midrule
\multirow{2}{*}{\textbf{\begin{tabular}[c]{@{}l@{}}Quality \\ Evaluation (3)\end{tabular}}}     
&Thematic Analysis  & S58  \\ 
&Reflexive Monitoring	& S40, S53 \\\midrule
\multirow{2}{*}{\textbf{\begin{tabular}[c]{@{}l@{}}Performance \\Evaluation (5)\end{tabular}}}          
& Performance Efficiency Metrics  & S1, S8, S27, S30   \\ 
& User Experience (UX) Hypothesis Testing & S1, S31 \\     \midrule
\textbf{No Evaluation (18)} & & \begin{tabular}[c]{@{}l@{}}
S14, S15, S18, S2, S23, S25, S3, S33, S37, S4, S49,\\S50, S52, S55, S56, S57, S63, S65\end{tabular}   \\ \hline                                                
\end{tabular}%
}
\caption{The evaluation method in each study}
\label{results evaluation of the study}
\end{table}

18 studies (26.09\%) did not use their requirements to evaluate a software solution, while the remaining 51 (73.91\%)) did with their described requirements. Table ~\ref{results evaluation of the study} categorizes evaluation types: stakeholder requirements (50 studies, 72.46\%), domain-specific requirements evaluation (7 studies, 10.14\%), quality evaluation (3 studies, 4.35\%), and performance evaluation (5 studies, 7.25\%). Of the 47 studies using requirements in evaluations, 19 employed multiple methods, and 28 used a single method.

End-user evaluation was conducted in 42 studies. Two used existing user data, and 11 employed user requirements reviews. For example, study S6 involved real-world users testing functional tasks to assess older adults' expectations and evaluate a health management app prototype. 
Domain-specific evaluation includes 2 studies using \emph{domain testing} and 5 studies using \emph{domain analysis}. For example, study S32 employed the System Usability Scale (SUS) and Technology Acceptance Model (TAM) to evaluate a mHealth app for rewarding healthy behaviour in older adults. In S46, SE engineers and medical experts provided feedback on a medication assistant system, using patients' clinical and functional requirements for testing. 
Quality evaluation includes \emph{thematic analysis}, \emph{iterative Verification}, and \emph{reflexive monitoring}, each containing 1 study. Thematic analysis refers to identifying, analyzing, and interpreting patterns or themes within qualitative data collected during software development projects. Study S58 used thematic analysis to help developers and caregivers understand the perspectives of older adults with dementia involved in the project. 
Statistical method evaluation includes \emph{performance efficiency metrics} and \emph{user experience (UX) hypothesis Testing}. Performance efficiency metrics involve metrics that evaluate the performance of a system which encompass statistical significance. For example, study S8 proposed a framework of underlying senior citizens' needs in smart-home services, A Balanced Scorecard (BSC) is used in study S8 to evaluate the effectiveness of services provided to the ageing population used evaluation, highlighting the statistical significance of the technology-supported ageing system in S8 for aged people. User experience (UX) hypothesis testing is a statistical test to determine whether a design leads to a higher user satisfaction score. 
\\
\begin{center}
\begin{myframe}[\centering\textbf{RQ2 Answer Summary}]
\footnotesize
The majority of our primary studies applied classic RE techniques, including interviews, surveys, workshops, and observations. Diverse models were used for RE in the studies. The most common ones are use cases, user stories, and personas. Few studies applied models like ontology models and conceptual models. Different tools were used for requirements elicitation and documentation, including popular prototyping tools (eg. Balsamiq or Axure RP) and UML modelling tools (eg. Enterprise Architect or Lucidchart) and less common ones like text data analysis software (eg. NVivo) and Application Lifecycle Management tools 
(eg. Tuleap). 
More than 80\% of studies validated their requirements, the most common way to validate it is to include end users in their design process. 75\% of the primary studies described the use of their requirements in diverse SE stages, and more than half of the studies have their development stage. Generally, the earlier a SE stage is, the more studies have used requirements in that stage. More than half of studies evaluated their solution with requirements, the most common one using workshops with end users, their caregivers, and technical experts. 
\end{myframe}
\end{center}

\subsection{RQ3 - What are the key Strengths, Limitations, Gaps, and Future work recommendations in the selected studies?}
\subsubsection{RQ3.1 - What are the key strengths/positive outcomes reported?}
\begin{table}[h!]
\centering
\resizebox{\textwidth}{!}{%
\begin{tabular}{lll}
Major Categories  & Minor Categories & Studies \\ \hline
\multirow{6}{*}{\textbf{\begin{tabular}[c]{@{}l@{}}User-centred \\ Approach\\ Enhancement (59)\end{tabular}}}
& \multirow{2}{*}{\begin{tabular}[c]{@{}l@{}}Functional Needs Understanding\end{tabular}}& \begin{tabular}[c]{@{}l@{}}
S1, S11, S12, S15, S16, S19, S22, S23, S27, S28, S30, \\ 
S31, S32, S33, S34, S35, S37, S42, S46, S47, S54, S55, \\ 
S58, S63, S7, S9, S39, S21, S69, S29, S67, S51, S66\end{tabular} \\
& \multirow{2}{*}{\begin{tabular}[c]{@{}l@{}}Participatory Design \\ Process Enhancement\end{tabular}}& \begin{tabular}[c]{@{}l@{}}
S24, S27, S28, S30, S34, S35, S37, S38, S42, S44, S46, \\ 
S48, S49, S5, S53, S56, S57, S60, S61, S63, S65, S7, \\
S68, S59, S66\end{tabular}                                             \\
& \multirow{2}{*}{\begin{tabular}[c]{@{}l@{}}Personalisation \& Adaptive Needs\end{tabular}} & \begin{tabular}[c]{@{}l@{}}
S2, S11, S12, S16, S18, S2, S41, S42, S45, S46, S47, \\ 
S49, S50, S53, S54, S57, S58, S8, S29, S66\end{tabular} \\
& Mutual Learning Promoting & \begin{tabular}[c]{@{}l@{}}
S11, S12, S26, S28, S37, S38, S4, S57, S61, S63, S68, \\
S59, S51\end{tabular} \\
& Clinical Goals Understanding  & 
S13, S25, S34, S38, S42, S50, S60, S29\\
& Emotional Needs Identifying  & S3, S4, S41, S54, S64 \\ \hline
\multirow{2}{*}{\textbf{\begin{tabular}[c]{@{}l@{}}Automation/AI/ML\\ Enhancement (34)\end{tabular}}}& AI/ML based System Improvement & \begin{tabular}[c]{@{}l@{}}
S12, S14, S15, S16, S17, S19, S18, S2, S20, S25, S26, \\ 
S30, S31, S35, S55, S7, S37, S41, S45, S46, S5, S52, \\ 
S56, S60, S61, S62, S63, S64, S9, S68, S21, S69, S51\end{tabular}  \\
& Saving time and effort by Automation & S27, S35, S37, S66 \\ \hline
\multirow{2}{*}{\textbf{\begin{tabular}[c]{@{}l@{}}User Health \\ Improvement (15)\end{tabular}}} & Maintaining Physical and Mental Health& 
S17, S19, S36, S37 \\
& Changing Lifestyle for Health Improvement& \begin{tabular}[c]{@{}l@{}}
S17, S19, S18, S24, S26, S30, S36, S47, S49, S50, \\S62, S29, S59, S66\end{tabular} \\ \hline
\multirow{4}{*}{\textbf{\begin{tabular}[c]{@{}l@{}}System-Building Process \\ Improvement (29)\end{tabular}}} 
& Improving/Streamlining Development & \begin{tabular}[c]{@{}l@{}}
S1, S14, S2, S48, S55, S57, S6, S60, S61, S62, S64, \\ S68, S29, S59\end{tabular} \\
& Enhancing Technology Adoption & S1, S22, S26, S27, S28, S3, S52, S6, S60, S62, S39 \\
& Enhance User Testing \& Feedback & S1, S32, S34, S38, S40, S42, S64, S8, S9 \\
& Improving Usability   & S1, S28, S32, S34, S36, S62, S39, S68 \\ \hline
\multirow{2}{*}{\textbf{\begin{tabular}[c]{@{}l@{}} Refinement of RE\\ Methods\&Guidelines (9) \end{tabular}}} 
&Existing RE methods Improvement & S10, S25, S30, S31, S38, S8, S67  \\
& \begin{tabular}[c]{@{}l@{}}Guidelines Improvement\end{tabular} 
& S43, S45 \\ \hline
\end{tabular}
}
\caption{The key reported benefits/positive outcomes}
\label{key outcomes of the study}
\end{table}

Table ~\ref{key outcomes of the study} summarises the key benefits of the RE approaches, which include enhancing user-centred approaches (59 studies, 85.51\%), enhancement of automation/AI/ML approaches (34 studies, 49.28\%), improving user's health (15 studies, 21.74\%), system-building process improvement (29 studies, 42.03\%), and refinement of existing RE methods/guidelines (9 studies, 13.04\%). Of the 69 studies, 59 reported multiple benefits, while 9 highlighted only one. 

Enhancement of User-centred approaches includes \emph{functional needs understanding} (33 studies), \emph{participatory design process enhancement} (25 studies), \emph{personalisation tasks and adaptive needs} (19 studies), \emph{mutual learning promoting} (13 studies), \emph{clinical goals understanding} (8 studies), and \emph{emotional needs identifying} (5 studies). For instance, study S64 found that their patient-centered medication system effectively supported end-user needs including forgetfulness support, medication images, and the expiration date, while study S48 showed how co-design improved personalized experiences and stakeholder involvement. Study S49 revealed participatory design with older adults enhanced personalized health tool development. 

Enhancement of Automation/AI/ML approaches includes sub-categories of \emph{AI/ML-based system improvement} (33 studies) and \emph{time-saving automation} (4 studies). In aged care software systems, AI/ML-based systems are very common when it is related to smart places, AI assistants, and accident prediction. For example, in study S45, the RE process helped identify age-related preferences for voice assistants, which are used to develop guidelines and design voice user interfaces that cater to the needs and preferences of older users. 

User's health improvement includes \emph{maintaining physical and mental health} (4 studies), and \emph{changing lifestyle for health improvement} (14 studies). For example, the RE approaches used in study S37 contributed to the development of an integrated assistive technology system called Rosetta that is user-centred, customizable for people with dementia, providing information, monitoring daily activities, and encouraging mental health rehabilitation. 

System-building process improvement includes \emph{improving/streamlining development} (14 studies), \emph{enhancing technology adoption} (11 studies), \emph{user testing enhancement} (9 studies), and \emph{improving usability} (8 studies). For example, study S14 is one of the first attempts at applying features of Judgement Call in a new context. It was successful as the Judgement Call methodology has found great success at Microsoft and its affiliated partners but has only been used among product designers. Study S26 reports that the RE approaches used, specifically combining Ambient Intelligence (AmI) technology with user-centred design methods, can greatly increase the acceptance of intelligent systems. This approach aims to provide a better quality of life for elderly and disabled individuals, creating a safe and intuitive environment to facilitate household tasks and preserve their independence for a longer period. 

Refinement of existing RE methods/guidelines includes \emph{existing RE methods improvement} (7 studies) and \emph{guidelines improvement} (2 studies). For example, in study S67, RE approaches were used to improve an existing requirements modelling approach, combining the strengths of Volere templates, Use Cases, and SysML Requirements diagrams to increase the efficiency and quality of design and implementation results. In study 

\subsubsection{RQ3.2 - What are the key limitations reported?}
\begin{table}[h!]
\centering
\resizebox{\textwidth}{!}{%
\begin{tabular}{llll}
Major Categories                  & Key Limitations  & Reported in studies  & We identified in studies                                                                                                                      \\ \hline
\multirow{4}{*}{\textbf{SE Challenges (23)}}    
& \begin{tabular}[c]{@{}l@{}}System Usability\end{tabular} & \begin{tabular}[c]{@{}l@{}}S15, S23, S24, S30, S32, S36, \\S5, S6, S50, S52, S60, S61, \\S62, S8, S39, S29, S21\end{tabular} &    \\
& Implementation  & S9, S68, S69, S51, S59& S18, S20, S42, S46, S65 \\
& Personalization Absence & S64& S44 \\ \hline
\multirow{3}{*}{\textbf{RE Limitations (20)}}   
& \begin{tabular}[c]{@{}l@{}}Representation of\\Participants Bias\end{tabular}                       & \begin{tabular}[c]{@{}l@{}}S1, S10, S3, S34, S35, S37, \\S38, S40, S49, S50, S53\end{tabular}&S16, S31, S33\\
& RE Process Limitations& S13, S14, S25, S56, S59, S63, S67 &S2, S26\\
& Participants Engagement & S10, S11, S13, S22 & S19, S48, S57\\ \hline
\textbf{Human Aspect Inclusion (5)} & & S10, S28, S43, S58&S54 \\ \hline
\multirow{3}{*}{\textbf{Needs Challenges (9)}} 
& Healthcare Needs & S27, S47, S55, S66&S12\\
& User Emotional Concerns& S41, S45, S7, S59, S67& S4, S65\\
& User Needs Conflicts& S45, S7&\\ \hline
\end{tabular}
}
\caption{The key reported limitations }
\label{key limitations of the study}
\end{table}

Table ~\ref{key limitations of the study} summarises the key limitations reported with the RE approaches used in the selected primary studies. The main issues reported (with overlaps) include SE challenges (23 studies, 33.33\%), RE methodological limitations (20 studies, 28.99\%), human aspect inclusion (5 studies, 7.25\%), and needs challenges (9 studies, 13.04\%). Of 52 studies that reported their key limitations, 4 presented more than one limitation, while the remaining 48 reported one key limitation. Studies S12, S16, S17, S19, S18, S2, S4, S20, S26, S31, S33, S42, S44, S46, S48, S57, and S65 did not report specific limitations, but we identified some limitations that we found in these studies in the last column in table ~\ref{key limitations of the study}.
SE challenges include \emph{usability/adoption/complexity of system functionalities}, \emph{implementation challenges}, and \emph{lack of AI-based personalized functions}. For example, in study S15, the authors noted that their Independent Living Support (ILS) systems could be enhanced with functionalities catering to users with different skill levels, collaborative interactions, and privacy controls. In study S9, the pattern recognition implementation was hindered by the lack of real user data needed for advanced algorithms, affecting RE extraction.
RE methodological limitations include \emph{representation and bias of participants}, \emph{RE process limitations}, and \emph{participants' engagement}. For example, in study S3, the survey questionnaire may not fully represent Indonesia's diverse regions, and participants might struggle to identify their specific ageing group due to the country's cultural diversity. 
Human aspect inclusion may overlook factors like gender, education level, socio-economic status, and culture/ethnicity. For instance, study S54 shows that seniors can easily interact with the CogniPlay platform, which accounts for age-related needs. However, aspects like motor skill and cognitive decline, perceptual changes, and limitations in spatial cognition and language comprehension should also be considered.
Needs challenges include \emph{healthcare needs challenges}, \emph{users needs conflicts}, and \emph{user emotional concerns}. In study S27, the authors reported challenges in balancing technical and clinical requirements, emphasizing the need for effective communication between technical and clinical teams. Study S41 highlights threats to validity related to emotional goals, stressing the importance of addressing emotional concerns in smart home technologies. 

\subsubsection{RQ3.3 - What are key recommendations for future research?}
\begin{table}[h!]
\centering
\resizebox{\textwidth}{!}{%
\begin{tabular}{llll}
Categories &
  \textbf{sub-Categories} &
  \textbf{Papers} &
  \textbf{Percentage} \\ \hline
\multirow{3}{*}{\textbf{System Advancement}} &
  Further Development &
  \begin{tabular}[c]{@{}l@{}}S1, S17, S18, S23, S35, S36, S38, S42, \\ S46, S47, S50, S52, S53, S54, S55, S56, \\ S64, S65, S21, S29\end{tabular} &
  \multirow{3}{*}{44.93\%} \\
 &
  Longitudinal Evaluation &
  \begin{tabular}[c]{@{}l@{}}S11, S24, S28, S34, S35, S43, S44, S50, \\ S60, S8, S39, S69, S59\end{tabular} &
   \\ \hline
\multirow{3}{*}{\textbf{RE Enhancement}} &
  Co-design &
  \begin{tabular}[c]{@{}l@{}}S19, S20, S25, S28, S3, S48, S49, S50, \\ S57, S63, S9, S39, S68, S51\end{tabular} &
  \multirow{3}{*}{30.43\%} \\
 &
  Participant Representation and Bias &
  S14, S24, S30, S31, S32, S37, S58, S63 &
   \\
 &
  RE models improvement &
  S10, S13, S16, S26, S7 &
   \\ \hline
\textbf{Guideline Improvement} &
  Improve WBAH Definition &
  S12 &
  1.45\% \\ \hline
\multirow{7}{*}{\textbf{Focus on Diverse Needs}} &
  Personalised Tasks &
  \begin{tabular}[c]{@{}l@{}}S2, S22, S27, S40, S5, S6, S61, S62, S67, \\ S51, S66\end{tabular} &
   \\
   &
  Social-technical requirements &
  S15 &
  \multirow{5}{*}{23.19\%} \\
 &
  Emotional Goals &
  S41 &
   \\
 &
  Loneliness &
  S4 &
   \\
 &
  Psychological Factors &
  S33 &
   \\
 &
  Privacy Protections &
  S45 &
   \\ \hline

\end{tabular}%
}
\caption{The key recommendations for future research}
\label{Key future work recommendations of the study}
\end{table}

Table ~\ref{Key future work recommendations of the study} summarises the key future work that is reported in each primary study. 13 studies did not report any future work recommendations, including S17, S19, S18, S31, S38, S4, S42, S44, S48, S54, S57, S65, S51. 

\paragraph{System Advancement}
Approximately 45\% of the studies reviewed have emphasized the importance of advancing the state of current system design using better requirements. Studies S38 and S55 specifically highlighted the potential benefits of further developing the system into a multi-platform solution to accommodate more diverse users and adding functionalities such as recovery and early mobility information. Several studies (eg. S11, S28, S34) have identified inadequacies in using longitudinal studies for understanding the evolving behaviours and needs of senior end users over time. 
Future research should examine system adaptability for seniors with diverse needs, including vision, hearing, mobility impairments, pain, and dementia. 

\paragraph{RE Enhancement \& Guideline Improvement}

Fourteen studies suggest improving co-design to better understand true needs, and real-time adaptive telecare functionalities, and enhance the interoperability of software agents. Study S50 specifically focuses on addressing the needs of different generations of older adults. Participant bias has been highlighted by eight studies, including S58, which proposes the use of empathy to understand different groups in participatory design, and S30, which recommends improving the balance of participant data to consider a comprehensive user group in RE. Five studies have mentioned the need for RE model improvements, such as incorporating user stories for non-English speakers as noted in study S7. Additionally, study S12 emphasizes the need for guideline improvement to understand better the trade-offs between needs, technical limitations, and costs.
 Future studies should aim to: 1) evaluate the effectiveness of formalized RE processes in software development; 2) propose guidelines for RE in agent-based digital health systems for disease diagnosis, health monitoring, and promoting healthy fitness and lifestyle; and 3) explore the dynamics of relationships through qualitative research, including interviews and focus groups with senior users and their caregivers or clinicians. Defining, understanding, and modelling these conflicts can guide the design of technologies that support not only the functional and clinical needs but also the emotional and social well-being of senior users. The relationships between users, particularly caregivers and families, and their influence on senior users are also of interest. 

\paragraph{Focus on Diverse Human Needs}
Incorporating diverse human aspects is crucial for effective digital health software for older adults. Classic studies have highlighted the importance of socio-technical factors, including emotional and psychological impacts. Eleven studies, including S6 and S67, suggested that adding personalized tips, such as drug information, meals, beauty tips, and clinical goals, can be beneficial. Studies S15, S41, S4, S33, and S45 each proposed diverse needs that should be better considered in future work. Additionally, the current studies that have applied human aspects have left a significant space for future exploration due to the limited types of human aspects they adopted and their uneven application across different studies. 
Future research should thus focus on 1) further development of human aspects in aged care digital health software, especially for clinical goals, emotional goals, psychological factors, and loneliness. Human aspects such as culture, age, language, mild mental health conditions, and education level significantly influence technology adoption and usage. They should also 2) investigate different methods for capturing these diverse factors and integrating them into the design of healthcare technologies. For instance, cultural considerations can impact the acceptance of smart home devices, while language and educational differences can affect the usability of health management applications. 3) utilize advanced statistical methods and machine learning algorithms to analyze the data, identifying trends and patterns that can inform the development of adaptive and responsive AI solutions.

\paragraph{Data-driven RE}
A further suggestion from our synthesis of current study gaps is to focus on data-driven RE modelling in developing adaptive personas and RE models that are specifically tailored to local and target aged care groups. This involves addressing 2 main research topics: 1) generating personas and RE models based on published studies and making them adaptive for specific aged care groups, and 2) the understanding and development of the data-driven personalized needs of senior users.
To generate personas and RE models based on published studies, future studies should explore methodologies for mining existing literature and data sources to construct initial personas. These personas should then be validated and adapted through local user studies, ensuring that they accurately reflect the needs and behaviours of the target population. 

\paragraph{Data-driven Health Solutions}
Understanding the personalized needs of senior users requires comprehensive data collection and analysis. Future research should focus on identifying key data sources, such as electronic health records (EHRs), surveys, and observational studies, to capture a wide range of user preferences and requirements. Specific areas of interest could include personalized meals and beauty tips, which can significantly enhance the quality of life for senior users. By leveraging data analytics and user feedback, researchers can develop tailored recommendations that meet the unique needs of individuals.


\paragraph{Integrated solutions for multiple health and wellness challenges}
Many proposed solutions target a single or small number of health and wellness challenges that ageing people face. This means multiple, sometimes incompatible, and very differently designed and realised solutions result from multiple RE exercises. Future work should explore capturing diverse, related health and wellness challenges ageing people face and the requirements for integrated, holistic supporting applications.  These will need personalisation to diverse human aspects of different people, including living conditions, background, physical and mental challenges, and preferences.

\begin{center}
\begin{myframe}[\centering\textbf{RQ3 Answer Summary}]
\footnotesize
Most of the selected primary studies mentioned the key positive outcomes of their studies in enhancing user-centred approach, providing automation/AI/ML, improving user health, and optimizing system development. A smaller number of studies refined existing RE methods and guidelines. More than one-quarter of the studies did not mention any key limitations of their studies. Among those mentioned, limitations included the need for further target system development, RE methodological limitations, lack of human aspect inclusion, and the difficulty of fulfilling diverse user needs. Some key future work recommendations include RE method improvements, new user engagement and representation methods, design and development recommendations, technology usability improvement, further human aspect inclusion, data-driven RE and systems, and integrated holistic support systems.
\end{myframe}
\end{center}


\section{Threats to Validity}

\subsection{Internal Validity}
To mitigate internal validity threats, we developed a detailed SLR protocol, reviewed by all authors, which guided the search process. We optimized the search string across multiple databases, using the systematic literature review tool "Covidence" for several filtering rounds to minimize selection bias, starting from titles and abstracts to full paper reviews. All authors participated in pilot tests for consistent data extraction.

\subsection{Construct Validity}
To reduce threats to construct validity, we conducted a comprehensive search across eight relevant databases using automated and manual strategies. We refined inclusion and exclusion criteria through discussions to select the most relevant studies. We also addressed inconsistent terminology for "aged care" and "geriatric syndromes," as well as "needs" and "requirements," by searching across terms used by researchers from various fields, highlighting gaps between SE, digital health, and medicine.

\subsection{Conclusion Validity}
To enhance conclusion validity, we developed a data extraction form aligned with our RQs and conducted a pilot test. All authors performed data extraction for a subset of studies, and results were compared for consistency. Deliberative discussions ensured accurate categorization and presentation, reducing bias in our analysis.

\subsection{External Validity}
To address external validity concerns, we employed a systematic search approach, combining automated and manual searches following established guidelines ~\cite{kitchenham2015evidence, kitchenham2022segress}. Our inclusion criteria focused on peer-reviewed academic studies, without publication dates or quality restrictions, to minimize publication bias.


\section{Conclusion}
We conducted a systematic literature review of requirements engineering for digital health software targeting senior users. After searching and filtering, we identified 69 primary studies. We identified various types of studies, health and well-being issues, a range of senior participants, and human aspects studied. We analysed the RE techniques, requirements modelling and validation approaches, and whether the requirements were used to build a system. We identify a range of strengths, limitations and gaps in the primary studies and potential future research directions.

\section{Declaration of generative AI and AI-assisted technologies in the writing process}
During preparation, Yuqing Xiao used ChatGPT and Grammarly for grammar correction, then reviewed and edited the content and takes full responsibility for the content of the published article.

\FloatBarrier

 \bibliographystyle{elsarticle-num} 
 \bibliography{elsarticle-template-num}

\begin{thebibliography}{100}
\expandafter\ifx\csname url\endcsname\relax
  \def\url#1{\texttt{#1}}\fi
\expandafter\ifx\csname urlprefix\endcsname\relax\def\urlprefix{URL }\fi
\expandafter\ifx\csname href\endcsname\relax
  \def\href#1#2{#2} \def\path#1{#1}\fi

\bibitem{un2022world}
World population prospects 2022, \url{https://www.un.org/development/desa/pd/sites/www.un.org.development.desa.pd/files/wpp2022_summary_of_results.pdf}, accessed: 2022-05-16.

\bibitem{un_ageing}
{United Nations}, \href{https://www.un.org/en/global-issues/ageing}{Ageing}.
\newline\urlprefix\url{https://www.un.org/en/global-issues/ageing}

\bibitem{brownie2013effects}
S.~Brownie, S.~Nancarrow, Effects of person-centered care on residents and staff in aged-care facilities: a systematic review, Clinical interventions in Aging (2013) 1--10.

\bibitem{grundy2023vision}
J.~Grundy, A.~Madugalla, J.~McIntosh, T.~Tran, Vision: Requirements engineering for software development in aged care, in: 2023 IEEE 31st International Requirements Engineering Conference Workshops (REW), IEEE, 2023, pp. 440--445.

\bibitem{jiang2020machine}
D.~Jiang, J.~Liao, H.~Duan, Q.~Wu, G.~Owen, C.~Shu, L.~Chen, Y.~He, Z.~Wu, D.~He, et~al., A machine learning-based prognostic predictor for stage iii colon cancer, Scientific reports 10~(1) (2020) 10333.

\bibitem{hyland2020early}
S.~L. Hyland, M.~Faltys, M.~H{\"u}ser, X.~Lyu, T.~Gumbsch, C.~Esteban, C.~Bock, M.~Horn, M.~Moor, B.~Rieck, et~al., Early prediction of circulatory failure in the intensive care unit using machine learning, Nature medicine 26~(3) (2020) 364--373.

\bibitem{subramanian2020precision}
M.~Subramanian, A.~Wojtusciszyn, L.~Favre, S.~Boughorbel, J.~Shan, K.~B. Letaief, N.~Pitteloud, L.~Chouchane, Precision medicine in the era of artificial intelligence: implications in chronic disease management, Journal of translational medicine 18 (2020) 1--12.

\bibitem{laplante2022requirements}
P.~A. Laplante, M.~Kassab, Requirements engineering for software and systems, Auerbach Publications, 2022.

\bibitem{karolita2023use}
D.~Karolita, J.~McIntosh, T.~Kanij, J.~Grundy, H.~O. Obie, Use of personas in requirements engineering: A systematic mapping study, Information and Software Technology (2023) 107264.

\bibitem{dalpiaz2020conceptualizing}
F.~Dalpiaz, A.~Sturm, Conceptualizing requirements using user stories and use cases: a controlled experiment, in: International Working Conference on Requirements Engineering: Foundation for Software Quality, Springer, 2020, pp. 221--238.

\bibitem{belani_towards_2022}
H.~Belani, P.~Šolić, T.~Perković, Towards {Ontology}-{Based} {Requirements} {Engineering} for {IoT}-{Supported} {Well}-{Being}, {Aging} and {Health}, in: 2022 {IEEE} 30th {International} {Requirements} {Engineering} {Conference} {Workshops} ({REW}), 2022, pp. 65--74.

\bibitem{heyn2021requirement}
H.-M. Heyn, E.~Knauss, A.~P. Muhammad, O.~Eriksson, J.~Linder, P.~Subbiah, S.~K. Pradhan, S.~Tungal, Requirement engineering challenges for ai-intense systems development, in: 2021 IEEE/ACM 1st Workshop on AI Engineering-Software Engineering for AI (WAIN), IEEE, 2021, pp. 89--96.

\bibitem{nazir2017applications}
F.~Nazir, W.~H. Butt, M.~W. Anwar, M.~A. Khan~Khattak, The applications of natural language processing (nlp) for software requirement engineering-a systematic literature review, Information Science and Applications 2017: ICISA 2017 8 (2017) 485--493.

\bibitem{ahmad2023requirements}
K.~Ahmad, M.~Abdelrazek, C.~Arora, M.~Bano, J.~Grundy, Requirements engineering for artificial intelligence systems: A systematic mapping study, Information and Software Technology 158 (2023) 107176.

\bibitem{billings1995approaches}
J.~R. Billings, S.~Cowley, Approaches to community needs assessment: a literature review, Journal of Advanced Nursing 22~(4) (1995) 721--730.

\bibitem{watkins1998needs}
R.~Watkins, D.~Leigh, W.~Platt, R.~Kaufman, Needs assessment—a digest, review, and comparison of needs assessment literature, performance improvement 37~(7) (1998) 40--53.

\bibitem{sleezer2014practical}
C.~M. Sleezer, D.~F. Russ-Eft, K.~Gupta, A practical guide to needs assessment, John Wiley \& Sons, 2014.

\bibitem{sharma2018using}
A.~Sharma, R.~A. Harrington, M.~B. McClellan, M.~P. Turakhia, Z.~J. Eapen, S.~Steinhubl, J.~R. Mault, M.~D. Majmudar, L.~Roessig, K.~J. Chandross, et~al., Using digital health technology to better generate evidence and deliver evidence-based care, Journal of the American College of Cardiology 71~(23) (2018) 2680--2690.

\bibitem{thomas2014review}
J.~G. Thomas, D.~S. Bond, Review of innovations in digital health technology to promote weight control, Current diabetes reports 14 (2014) 1--10.

\bibitem{batra2017digital}
S.~Batra, R.~A. Baker, T.~Wang, F.~Forma, F.~DiBiasi, T.~Peters-Strickland, Digital health technology for use in patients with serious mental illness: a systematic review of the literature, Medical Devices: Evidence and Research (2017) 237--251.

\bibitem{pal2017smart}
D.~Pal, T.~Triyason, S.~Funikul, Smart homes and quality of life for the elderly: a systematic review, in: 2017 IEEE international symposium on multimedia (ISM), IEEE, 2017, pp. 413--419.

\bibitem{matayong2023iot}
S.~Matayong, K.~Jetwanna, C.~Choksuchat, S.~Choosawang, N.~Trakulmaykee, S.~Limsuwan, K.~Inthanuchit, Iot-based systems and applications for elderly healthcare: a systematic review, Universal Access in the Information Society (2023) 1--27.

\bibitem{cysneiros2002requirements}
L.~M. Cysneiros, Requirements engineering in the health care domain, in: Proceedings IEEE Joint International Conference on Requirements Engineering, IEEE, 2002, pp. 350--356.

\bibitem{levy2023sustaining}
M.~Levy, E.~C. Groen, K.~Taveter, D.~Amyot, E.~Yu, L.~Liu, I.~Richardson, M.~Spichkova, A.~Jussli, S.~Mosser, Sustaining human health: A requirements engineering perspective, Journal of Systems and Software 204 (2023) 111792.

\bibitem{aziz2016requirement}
M.~W. Aziz, A.~A. Sheikh, E.~A. Felemban, Requirement engineering technique for smart spaces, in: Proceedings of the International Conference on Internet of things and Cloud Computing, 2016, pp. 1--7.

\bibitem{ariaeinejad2016user}
M.~Ariaeinejad, N.~Archer, M.~Stacey, T.~Rapanos, F.~Elias, F.~Naji, User-centered requirements analysis and design solutions for chronic disease self-management, in: HCI in Business, Government, and Organizations: Information Systems: Third International Conference, HCIBGO 2016, Held as Part of HCI International 2016, Toronto, Canada, July 17-22, 2016, Proceedings, Part II 3, Springer, 2016, pp. 3--15.

\bibitem{fischer2020importance}
B.~Fischer, A.~Peine, B.~{\"O}stlund, The importance of user involvement: a systematic review of involving older users in technology design, The Gerontologist 60~(7) (2020) e513--e523.

\bibitem{merkel2019participatory}
S.~Merkel, A.~Kucharski, Participatory design in gerontechnology: a systematic literature review, The Gerontologist 59~(1) (2019) e16--e25.

\bibitem{chute2022user}
C.~Chute, T.~French, S.~Raman, J.~Bradley, User requirements for comanaged digital health and care, Journal of Medical Internet Research 24~(6) (2022) e35337.

\bibitem{zhang2023application}
A.~R. Zhang, S.~Attrill, J.~Eliott, R.~A. Ankeny, P.~Moynihan, Application of co-design in residential aged care: a scoping review protocol, JBI Evidence Synthesis 21~(8) (2023) 1665--1671.

\bibitem{page2021prisma}
M.~J. Page, J.~E. McKenzie, P.~M. Bossuyt, I.~Boutron, T.~C. Hoffmann, C.~D. Mulrow, L.~Shamseer, J.~M. Tetzlaff, E.~A. Akl, S.~E. Brennan, et~al., The prisma 2020 statement: an updated guideline for reporting systematic reviews, Bmj 372 (2021).

\bibitem{kitchenham2015evidence}
B.~A. Kitchenham, D.~Budgen, P.~Brereton, Evidence-based software engineering and systematic reviews, Vol.~4, CRC press, 2015.

\bibitem{richardson1995well}
W.~S. Richardson, M.~C. Wilson, J.~Nishikawa, R.~S. Hayward, The well-built clinical question: a key to evidence-based decisions, ACP journal club 123~(3) (1995) A12--A13.

\bibitem{wohlin2014guidelines}
C.~Wohlin, Guidelines for snowballing in systematic literature studies and a replication in software engineering, in: Proceedings of the 18th international conference on evaluation and assessment in software engineering, 2014, pp. 1--10.

\bibitem{mesko2017digital}
B.~Mesk{\'o}, Z.~Drobni, {\'E}.~B{\'e}nyei, B.~Gergely, Z.~Gy{\H{o}}rffy, Digital health is a cultural transformation of traditional healthcare, Mhealth 3 (2017).

\bibitem{muller2022genaichi}
M.~Muller, L.~B. Chilton, A.~Kantosalo, C.~P. Martin, G.~Walsh, Genaichi: generative ai and hci, in: CHI conference on human factors in computing systems extended abstracts, 2022, pp. 1--7.

\bibitem{alshamrani2022iot}
M.~Alshamrani, Iot and artificial intelligence implementations for remote healthcare monitoring systems: A survey, Journal of King Saud University-Computer and Information Sciences 34~(8) (2022) 4687--4701.

\bibitem{golomb2012older}
B.~A. Golomb, V.~T. Chan, M.~A. Evans, S.~Koperski, H.~L. White, M.~H. Criqui, The older the better: are elderly study participants more non-representative? a cross-sectional analysis of clinical trial and observational study samples, BMJ open 2~(6) (2012) e000833.

\bibitem{guralnik2000ratio}
J.~Guralnik, J.~Balfour, S.~Volpato, The ratio of older women to men: historical perspectives and cross-national comparisons, Aging Clinical and Experimental Research 12 (2000) 65--76.

\bibitem{jensen1994distribution}
R.~K. Jensen, P.~Fletcher, Distribution of mass to the segments of elderly males and females, Journal of biomechanics 27~(1) (1994) 89--96.

\bibitem{kawamoto2013application}
A.~L.~S. Kawamoto, V.~F. Martins, Application designed for the elderly using gestural interface, Revista Brasileira de Computa{\c{c}}{\~a}o Aplicada 5~(2) (2013) 96--109.

\bibitem{kawamoto2014converging}
A.~L.~S. Kawamoto, V.~F. Martins, F.~S.~C. da~Silva, Converging natural user interfaces guidelines and the design of applications for older adults, in: 2014 IEEE International Conference on Systems, Man, and Cybernetics (SMC), IEEE, 2014, pp. 2328--2334.

\bibitem{kitchenham2022segress}
B.~Kitchenham, L.~Madeyski, D.~Budgen, Segress: Software engineering guidelines for reporting secondary studies, IEEE Transactions on Software Engineering 49~(3) (2022) 1273--1298.

\bibitem{boontarig_factors_2012}
W.~Boontarig, W.~Chutimaskul, V.~Chongsuphajaisiddhi, B.~Papasratorn, Factors influencing the thai elderly intention to use smartphone for e-health services, in: 2012 IEEE symposium on humanities, science and engineering research, IEEE, 2012, pp. 479--483.

\bibitem{evans_requirements_2014}
C.~Evans, L.~Brodie, J.~C. Augusto, Requirements {Engineering} for {Intelligent} {Environments}, in: 2014 {International} {Conference} on {Intelligent} {Environments}, 2014, pp. 154--161.

\bibitem{santoso_berbakti_2016}
H.~A. Santoso, F.~Firdausillah, S.~E. Sukmana, A.~Yusriana, A.~Juliandri, T.~Witjahjono, Berbakti: An elderly apps for strengthen parent-children relationship in indonesia, in: 2016 International Seminar on Application for Technology of Information and Communication (ISemantic), IEEE, 2016, pp. 327--331.

\bibitem{mulvenna_participatory_2017}
M.~Mulvenna, H.~Zheng, R.~Bond, P.~McAllister, H.~Wang, R.~Riestra, Participatory design-based requirements elicitation involving people living with dementia towards a home-based platform to monitor emotional wellbeing, in: 2017 IEEE international conference on bioinformatics and biomedicine (BIBM), IEEE, 2017, pp. 2026--2030.

\bibitem{cahill_design_2017}
J.~Cahill, S.~McLoughlin, D.~Blazek, The design of new technologies addressing independence, social participation \& wellness for older people domicile in residential homes, in: 2017 International Conference on Computational Science and Computational Intelligence (CSCI), IEEE, 2017, pp. 1672--1677.

\bibitem{musthafa_towards_2020}
F.~N. Musthafa, M.~B. Mustafa, F.~P. Tajudeen, T.~S. Ramasamy, A.~Sinniah, Towards the development of a user-centred health management application for elderly, in: 2020 2nd International Conference on Advancements in Computing (ICAC), Vol.~1, IEEE, 2020, pp. 13--18.

\bibitem{abdullah_using_2020}
N.~N.~B. Abdullah, J.~Grundy, J.~McIntosh, Y.~C. How, S.~Saharuddin, K.~K. Tat, E.~ShinYe, A.~J.~A. Rastom, N.~L. Othman, Using {Work} {System} {Design}, {User} {Stories} and {Emotional} {Goal} {Modeling} for an {mHealth} {System}, in: 2020 {IEEE} {First} {International} {Workshop} on {Requirements} {Engineering} for {Well}-{Being}, {Aging}, and {Health} ({REWBAH}), 2020, pp. 1--10.

\bibitem{wei_understanding_2020}
Z.~Wei, Y.~Liu, L.~Liu, E.~Yu, J.~Mylopoulos, C.~K. Chang, Understanding {Requirements} for {Technology}-{Supported} {Healthy} {Aging}, in: 2020 {IEEE} {First} {International} {Workshop} on {Requirements} {Engineering} for {Well}-{Being}, {Aging}, and {Health} ({REWBAH}), 2020, pp. 47--56.

\bibitem{grave_requirement_2021}
A.~Grave, S.~Robben, M.~Oey, S.~B. Allouch, M.~Mohammadi, Requirement elicitation and prototype development of an intelligent environment to support people with early dementia, in: 2021 17th International Conference on Intelligent Environments (IE), IEEE, 2021, pp. 1--8.

\bibitem{jussli_senior_2021}
A.~Jussli, H.~Gewald, Senior {DT} - {A} {Design} {Thinking} {Method} to {Improve} {Requirements} {Engineering} for {Elderly} {Citizens}, in: 2021 {IEEE} 29th {International} {Requirements} {Engineering} {Conference} {Workshops} ({REW}), 2021, pp. 240--247.

\bibitem{rauer_eliciting_2021}
J.~R. Rauer, K.~Kolluri, L.~Chung, C.~Liu, T.~Hill, Eliciting {Smartphone} {App} {Requirements} for {Helping} {Senior} {People}: {A} {Questionnaire} {Approach}, in: 2021 {IEEE} 29th {International} {Requirements} {Engineering} {Conference} {Workshops} ({REW}), 2021, pp. 278--287.

\bibitem{radeck_understanding_2022}
L.~Radeck, B.~Paech, F.~Kramer-Gmeiner, M.~Wettstein, H.-W. Wahl, A.-L. Schubert, U.~Sperling, Understanding {IT}-related {Well}-being, {Aging} and {Health} {Needs} of {Older} {Adults} with {Crowd}-{Requirements} {Engineering}, in: 2022 {IEEE} 30th {International} {Requirements} {Engineering} {Conference} {Workshops} ({REW}), 2022, pp. 57--64.

\bibitem{robinson_participatory_2022}
K.-M. Robinson, R.~Devkota, J.~Millar, A {Participatory} {Design} {Methodology} to {Elicit} {Aging}- in-{Place} {Stakeholder} {Concerns} with {Ambient} {Assistive} {Living} ({AAL}) {Devices} {During} {COVID}-19, in: 2022 {IEEE} 30th {International} {Requirements} {Engineering} {Conference} {Workshops} ({REW}), 2022, pp. 38--47.

\bibitem{bella_challenges_2014}
G.~Bella, P.~Jappinen, J.~Laakkonen, The {Challenges} behind {Independent} {Living} {Support} {Systems}, in: Active {Media} {Technology}. 10th {International} {Conference}, {AMT} 2014. {Proceedings}: {LNCS} 8610, 2014, pp. 464 -- 74.

\bibitem{koshima_model-driven_2016}
A.~Koshima, V.~Englebert, M.~Amani, A.~Debieche, A.~Wakjira, A {Model}-{Driven} {Engineering} {Approach} for the {Well}-{Being} of {Ageing} {People}, in: Advances in {Conceptual} {Modeling}. {ER} 2016 {Workshops} {AHA}, {MoBiD}, {MORE}-{BI}, {MReBA}, {QMMQ}, {SCME} and {WM2SP}. {Proceedings}: {LNCS} 9975, 2016, pp. 21 -- 9.

\bibitem{goonetilleke_enhancing_2020}
R.~S. Goonetilleke, E.~Y.~L. Au, Enhancing the {Life} of the {Elderly} - {An} {Application} of {Design} {Thinking}, in: Advances in {Physical} {Ergonomics} and {Human} {Factors}, Springer International Publishing, 2020, pp. 388--396.

\bibitem{cleland_contextualizing_2015}
A.~C. Rodriguez, C.~Roda, P.~González, E.~Navarro, Contextualizing {Tasks} in {Tele}-{Rehabilitation} {Systems} for {Older} {People}, in: Ambient {Assisted} {Living}. {ICT}-based {Solutions} in {Real} {Life} {Situations}, Springer International Publishing, 2015, pp. 29--41.

\bibitem{beltran_smart_2022}
J.~Beltrán, O.~A. Montoya-Valdivia, R.~B.-D.~L. Torre, L.~Melendez-Lineros, G.~Parada-Picos, C.~B. Pérez, C.~Martínez-García-Moreno, Smart {Technologies} for {Monitoring} {Older} {Adults} with {Dementia}, in: Smart {Cities}, Springer International Publishing, 2022, pp. 116--127.

\bibitem{faria_active_2022}
G.~Faria, T.~Silva, J.~Abreu, Active and {Healthy} {Aging}: {The} {Role} of a {Proactive} {Information} {Assistant} {Embedded} on {TV}, in: Applications and {Usability} of {Interactive} {TV}, Springer Nature Switzerland, 2022, pp. 70--84.

\bibitem{le_design_2014}
T.~Le, B.~Reeder, J.~Chung, H.~Thompson, G.~Demiris, Design of smart home sensor visualizations for older adults, Technology and health care : official journal of the European Society for Engineering and Medicine (2014).

\bibitem{maria_luisa_rodriguez-almendros_design_2021}
M.~J. H. J. S.-J. C. R.-D. María Luisa Rodríguez-Almendros, María José Rodríguez-Fórtiz, S.~Rute-Pérez, Design guide and usability questionnaire to develop and assess {VIRTRAEL}, a web-based cognitive training tool for the elderly, Behaviour \& Information Technology (2021) 1355--1374.

\bibitem{sokoler_embracing_2007}
T.~Sokoler, M.~S. Svensson, Embracing ambiguity in the design of non-stigmatizing digital technology for social interaction among senior citizens, Behaviour \& Information Technology (2007) 297--307.

\bibitem{hazwani_mohd_mohadis_designing_2016}
N.~M.~A. Hazwani Mohd~Mohadis, A.~F. Smeaton, Designing a persuasive physical activity application for older workers: understanding end-user perceptions, Behaviour \& Information Technology (2016) 1102--1114.

\bibitem{kuusik_home_2012}
A.~Kuusik, E.~Reilent, K.~Sarna, M.~Parve, Home telecare and rehabilitation system with aspect oriented functional integration, Biomedizinische Technik (2012) 33 -- 6.

\bibitem{casas_user_2008}
R.~Casas, R.~Blasco~Marín, A.~Robinet, A.~R. Delgado, A.~R. Yarza, J.~McGinn, R.~Picking, V.~Grout, User {Modelling} in {Ambient} {Intelligence} for {Elderly} and {Disabled} {People}, in: Computers {Helping} {People} with {Special} {Needs}, Springer, 2008, pp. 114--122.

\bibitem{scandurra_user_2008}
I.~Scandurra, M.~Hägglund, S.~Koch, From user needs to system specifications: {Multi}-disciplinary thematic seminars as a collaborative design method for development of health information systems, Journal of Biomedical Informatics (2008) 557 -- 569.

\bibitem{berge_acceptability_2022}
L.~I. Berge, M.~H. Gedde, J.~C. Torrado~Vidal, B.~Husebo, K.~M. Hynninen, S.~E. Knardal, K.~G. Madsø, The acceptability, adoption, and feasibility of a music application developed using participatory design for home-dwelling persons with dementia and their caregivers. {The} “{Alight}” app in the {LIVE}@{Home}.{Path} trial, Frontiers in Psychiatry (2022).

\bibitem{willard_development_2018}
S.~Willard, G.~Cremers, Y.~P. Man, E.~v. Rossum, M.~Spreeuwenberg, L.~d. Witte, Development and testing of an online community care platform for frail older adults in the {Netherlands} (2018).

\bibitem{pal_internet--things_2018}
D.~Pal, S.~Funilkul, N.~Charoenkitkarn, P.~Kanthamanon, Internet-of-{Things} and {Smart} {Homes} for {Elderly} {Healthcare}: {An} {End} {User} {Perspective}, IEEE Access (2018) 10483--10496.

\bibitem{pal_embracing_2019}
D.~Pal, B.~Papasratorn, W.~Chutimaskul, S.~Funilkul, Embracing the {Smart}-{Home} {Revolution} in {Asia} by the {Elderly}: {An} {End}-{User} {Negative} {Perception} {Modeling}, IEEE Access (2019) 38535--38549.

\bibitem{ziefle_results_2020}
S.~Jansen-Kosterink, R.~Bulthuis, S.~Ter~Stal, L.~Van~Velsen, A.~Pnevmatikakis, S.~Kyriazakos, A.~Pomazanskyi, H.~Op~Den~Akker, The {Results} of an {Iterative} {Evaluation} {Process} of an {Mhealth} {Application} for {Rewarding} {Healthy} {Behaviour} {Among} {Older} {Adults}, in: Information and {Communication} {Technologies} for {Ageing} {Well} and e-{Health}, Springer International Publishing, 2020, pp. 62--78.

\bibitem{macdonald_designing_2007}
A.~Macdonald, D.~Loudon, Designing data to be inclusive: enabling cross-disciplinary and participative processes, Interacción (2007) 217--223.

\bibitem{umm_e_mariya_shah_usability_2022}
T.~K.~C. Umm~e Mariya~Shah, Y.~Mehmood, A {Usability} {Evaluation} {Instrument} for {Pain} {Management} {Mobile} {Applications}: {An} {Elderly}’s {Perspective}, International Journal of Human–Computer Interaction (2022) 1--17.

\bibitem{schafer_survey-based_2019}
K.~Schäfer, P.~Rasche, C.~Bröhl, S.~Theis, L.~Barton, C.~Brandl, M.~Wille, V.~Nitsch, A.~Mertens, Survey-based personas for a target-group-specific consideration of elderly end users of information and communication systems in the {German} health-care sector, International Journal of Medical Informatics (2019).

\bibitem{van_t_klooster_virtual_2011}
J.-W. van~'t Klooster, B.-J. van Beijnum, P.~Pawar, K.~Sikkel, L.~Meertens, H.~Hermens, Virtual communities for elderly healthcare: user-based requirements elicitation, International Journal of Networking and Virtual Organisations (2011) 214 -- 32.

\bibitem{meiland_participation_2014}
F.~Meiland, B.~Hattink, T.~Overmars-Marx, M.~E. de~Boer, A.~Jedlitschka, P.~W.~G. Ebben, I.~I. N.~W. Stalpers-Croeze, S.~E. Flick, J.~van~der Leeuw, I.~Karkowski, R.-M. Dröes, {Rose-Marie Dröes}, Participation of end users in the design of assistive technology for people with mild to severe cognitive problems; {The} {European} {Rosetta} project, International Psychogeriatrics (2014) 769--779.

\bibitem{tran-nguyen_mobile_2022}
K.~Tran-Nguyen, C.~Berger, R.~Bennett, M.~Wall, S.~N. Morin, F.~Rajabiyazdi, Mobile {App} {Prototype} in {Older} {Adults} for {Postfracture} {Acute} {Pain} {Management}: {User}-{Centered} {Design} {Approach}, JMIR Aging (2022).

\bibitem{pater_addressing_2017}
J.~Pater, S.~Owens, S.~Farmer, E.~Mynatt, B.~Fain, Addressing medication adherence technology needs in an aging population, in: Proceedings of the 11th {EAI} {International} {Conference} on {Pervasive} {Computing} {Technologies} for {Healthcare}, ACM, 2017, pp. 58--67.

\bibitem{van_hoof_what_2016}
J.~van Hoof, S.~T.~M. Peek, E.~Wouters, K.~Luijkx, H.~J.~M. Vrijhoef, What it takes to successfully implement technology for aging in place: focus groups with stakeholders, Journal of Medical Internet Research (2016).

\bibitem{curumsing_emotion-oriented_2019}
M.~K. Curumsing, N.~Fernando, M.~Abdelrazek, R.~Vasa, K.~Mouzakis, J.~Grundy, Emotion-oriented requirements engineering: {A} case study in developing a smart home system for the elderly, Journal of Systems and Software (2019) 215--229.

\bibitem{ariaeinejad_user-centered_2016}
M.~Ariaeinejad, N.~Archer, M.~Stacey, T.~Rapanos, F.~Elias, F.~Naji, User-centered requirements analysis and design solutions for chronic disease self-management, Lecture Notes in Computer Science (including subseries Lecture Notes in Artificial Intelligence and Lecture Notes in Bioinformatics) (2016) 3 -- 15.

\bibitem{kawamoto_visuospatial_2017}
A.~L.~S. Kawamoto, V.~F. Martins, A visuospatial memory game for the elderly using gestural interface, Lecture Notes in Computer Science (including subseries Lecture Notes in Artificial Intelligence and Lecture Notes in Bioinformatics) (2017) 430 -- 443.

\bibitem{stamm_exergames_2019}
O.~Stamm, S.~Vorwerg, U.~Müller-Werdan, Exergames in {Augmented} {Reality} for {Older} {Adults} with {Hypertension}: {A} {Qualitative} {Study} {Exploring} {User} {Requirements}, Lecture Notes in Computer Science (including subseries Lecture Notes in Artificial Intelligence and Lecture Notes in Bioinformatics) (2019) 232 -- 244.

\bibitem{gollasch_age-related_2021}
D.~Gollasch, G.~Weber, Age-{Related} {Differences} in {Preferences} for {Using} {Voice} {Assistants}, in: Mensch und {Computer} 2021, ACM, 2021, pp. 156--167.

\bibitem{ferreira_elderly_2014}
F.~Ferreira, N.~Almeida, A.~F. Rosa, A.~Oliveira, A.~L. de~Abreu~Oliveira, J.~Casimiro, C.~M. da~Silva, S.~Silva, {António Teixeira}, {António Teixeira}, A.~L. Teixeira, Elderly {Centered} {Design} for {Interaction} – {The} {Case} of the {S4S} {Medication} {Assistant}, Procedia Computer Science (2014) 398--408.

\bibitem{breidenbach_development_2022}
M.~Breidenbach, F.~Hamiti, A.~Guluzade, N.~Heiba, Y.~Mohamad, C.~Velasco, B.~Herbeck~Belnap, D.~L{\"u}hmann, Development of a flexible and interoperable architecture to customize clinical solutions targeting the care of multimorbid patients, in: Proceedings of the 10th International Conference on Software Development and Technologies for Enhancing Accessibility and Fighting Info-exclusion, 2022, pp. 12--17.

\bibitem{mohamad_key_2022}
Y.~Mohamad, H.~Gappa, N.~Heiba, M.~Yuksel, C.~A. Velasco, M.~Gencturk, P.~Abizanda, G.~B. Laleci~Erturkmen, A.~Steinhoff, J.~Ayadi, T.~N. Arvanitis, B.~Ahmad, O.~Pournik, I.~Kyrou, G.~Despotou, T.~Robbins, S.~Lim Choi~Keung, K.~Le, W.~Schmidt-Barzynski, Key scenarios, {Use} {Cases} \& {Architecture} of an {E}-health {Homecare} {Instance}, in: Proceedings of the 10th {International} {Conference} on {Software} {Development} and {Technologies} for {Enhancing} {Accessibility} and {Fighting} {Info}-exclusion, ACM, 2022, pp. 25--30.

\bibitem{martin-hammond_engaging_2018}
A.~Martin-Hammond, S.~Vemireddy, K.~Rao, Engaging {Older} {Adults} in the {Participatory} {Design} of {Intelligent} {Health} {Search} {Tools}, in: Proceedings of the 12th {EAI} {International} {Conference} on {Pervasive} {Computing} {Technologies} for {Healthcare}, Association for Computing Machinery, 2018, pp. 280--284.

\bibitem{davidson_what_2013}
J.~L. Davidson, C.~Jensen, What health topics older adults want to track: a participatory design study, in: Proceedings of the 15th {International} {ACM} {SIGACCESS} {Conference} on {Computers} and {Accessibility}, Association for Computing Machinery, 2013, pp. 1--8.

\bibitem{lee_steps_2017}
H.~R. Lee, S.~Šabanović, W.-L. Chang, S.~Nagata, J.~Piatt, C.~Bennett, D.~Hakken, Steps {Toward} {Participatory} {Design} of {Social} {Robots}: {Mutual} {Learning} with {Older} {Adults} with {Depression}, in: Proceedings of the 2017 {ACM}/{IEEE} {International} {Conference} on {Human}-{Robot} {Interaction}, ACM, 2017, pp. 244--253.

\bibitem{jorge_adaptive_2001}
J.~A. Jorge, Adaptive tools for the elderly: new devices to cope with age-induced cognitive disabilities, in: Proceedings of the 2001 {EC}/{NSF} workshop on {Universal} accessibility of ubiquitous computing: providing for the elderly, ACM, 2001, pp. 66--70.

\bibitem{sorgalla_improving_2017}
J.~Sorgalla, P.~Schabsky, S.~Sachweh, M.~Grates, E.~Heite, Improving {Representativeness} in {Participatory} {Design} {Processes} with {Elderly}, in: Proceedings of the 2017 {CHI} {Conference} {Extended} {Abstracts} on {Human} {Factors} in {Computing} {Systems}, ACM, 2017, pp. 2107--2114.

\bibitem{vasconcelos_designing_2012}
A.~Vasconcelos, P.~A. Silva, J.~Caseiro, F.~Nunes, L.~F. Teixeira, Designing tablet-based games for seniors: the example of {CogniPlay}, a cognitive gaming platform, in: Proceedings of the 4th {International} {Conference} on {Fun} and {Games}, ACM, 2012, pp. 1--10.

\bibitem{costa_designing_2021}
L.~Costa, J.~Carneiro, M.~Temporao, Designing an {App} for {Nursing} {Homes} to {Clinical} {Users}, in: Proceedings of the 5th {International} {Conference} on {Medical} and {Health} {Informatics}, Association for Computing Machinery, 2021, pp. 150--157.

\bibitem{aziz_requirement_2016}
M.~W. Aziz, A.~A. Sheikh, E.~A. Felemban, Requirement {Engineering} {Technique} for {Smart} {Spaces}, in: Proceedings of the {International} {Conference} on {Internet} of things and {Cloud} {Computing}, ACM, 2016, pp. 1--7.

\bibitem{askenas_supporting_2020}
L.~Askenas, J.~Aidemark, Supporting elderly living longer at home: a framework for building a sustainable eco-system, in: Proceedings of the {International} {Conferences} on {Internet} {Technologies} \&amp; {Society} ({ITS} 2020) and {Sustainability} {Technology} and {Education} ({STE} 2020), 2020, pp. 18 -- 26.

\bibitem{lindsay_empathy_2012}
S.~Lindsay, K.~Brittain, D.~Jackson, C.~Ladha, K.~Ladha, P.~Olivier, Empathy, participatory design and people with dementia, in: Proceedings of the {SIGCHI} {Conference} on {Human} {Factors} in {Computing} {Systems}, ACM, 2012, pp. 521--530.

\bibitem{verhoeven_mobiles_2016}
F.~Verhoeven, A.~Cremers, M.~Schoone, J.~V. Dijk, Mobiles for mobility: {Participatory} design of a ‘{Happy} walker’ that stimulates mobility among older people, Gerontechnology (2016) 32--44.

\bibitem{wang_mobile_2016}
J.~Wang, {Jing Wang}, D.~D. Carroll, M.~Peck, S.~Myneni, Y.~Gong, Mobile and {Wearable} {Technology} {Needs} for {Aging} in {Place}: {Perspectives} from {Older} {Adults} and {Their} {Caregivers} and {Providers}., Studies in health technology and informatics (2016) 486--490.

\bibitem{compagna_limits_2015}
D.~Compagna, F.~Kohlbacher, The limits of participatory technology development: {The} case of service robots in care facilities for older people, Technological Forecasting and Social Change (2015) 19--31.

\bibitem{cahill_design_2018}
J.~Cahill, S.~McLoughlin, S.~Wetherall, The {Design} of {New} {Technology} {Supporting} {Wellbeing}, {Independence} and {Social} {Participation}, for {Older} {Adults} {Domiciled} in {Residential} {Homes} and/or {Assisted} {Living} {Communities}, Technologies (2018) 18 (33 pp.) --.

\bibitem{chen_understanding_2021}
C.~Chen, J.~G. Johnson, K.~Charles, A.~Lee, E.~T. Lifset, M.~Hogarth, A.~A. Moore, E.~Farcas, N.~Weibel, Understanding {Barriers} and {Design} {Opportunities} to {Improve} {Healthcare} and {QOL} for {Older} {Adults} through {Voice} {Assistants}, in: The 23rd {International} {ACM} {SIGACCESS} {Conference} on {Computers} and {Accessibility}, ACM, 2021, pp. 1--16.

\bibitem{teixeira_design_2017}
A.~Teixeira, F.~Ferreira, N.~Almeida, S.~Silva, A.~Rosa, J.~Casimiro~Pereira, D.~Vieira, Design and development of {Medication} {Assistant}: older adults centred design to go beyond simple medication reminders, Universal Access in the Information Society (2017) 545 -- 60.

\bibitem{hendriks_designing_2013}
N.~Hendriks, F.~Truyen, E.~Duval, Designing with {Dementia}: {Guidelines} for {Participatory} {Design} together with {Persons} with {Dementia} (2013) 649--666.

\bibitem{nielsen_user-innovated_2018}
A.~C. Nielsen, G.~S. Rotger, A.~M. Kanstrup, L.~A. Laplante, User-{Innovated} {eHealth} {Solutions} for {Service} {Delivery} to {Older} {Persons} {With} {Hearing} {Impairment}, American Journal of Audiology (2018) 403--416.

\bibitem{meacham_requirements_nodate}
S.~Meacham, K.~Phalp, Requirements engineering methods for an {Internet} of {Things} application: fall-detection for ambient assisted living.

\bibitem{duh_applications_2016}
E.~S. Duh, J.~Guna, M.~Pogačnik, J.~Sodnik, Applications of {Paper} and {Interactive} {Prototypes} in {Designing} {Telecare} {Services} for {Older} {Adults}, Journal of Medical Systems (2016) 92.

\bibitem{eftring_designing_2016}
H.~Eftring, S.~Frennert, Designing a social and assistive robot for seniors, Zeitschrift für Gerontologie und Geriatrie (2016).

\end{thebibliography}






\appendix

\section{List of Selected Primary Studies}
\nobibliography
a
\sloppy
\begin{footnotesize}
\begin{itemize}[label={}]
\item \textbf{S1} \bibentry{boontarig_factors_2012}
\item \textbf{S2} \bibentry{evans_requirements_2014}
\item \textbf{S3} \bibentry{santoso_berbakti_2016}
\item \textbf{S4} \bibentry{mulvenna_participatory_2017}
\item \textbf{S5} \bibentry{cahill_design_2017}
\item \textbf{S6} \bibentry{musthafa_towards_2020}
\item \textbf{S7} \bibentry{abdullah_using_2020}
\item \textbf{S8} \bibentry{wei_understanding_2020}
\item \textbf{S9} \bibentry{grave_requirement_2021}
\item \textbf{S10} \bibentry{jussli_senior_2021}
\item \textbf{S11} \bibentry{rauer_eliciting_2021}
\item \textbf{S12} \bibentry{radeck_understanding_2022}
\item \textbf{S13} \bibentry{belani_towards_2022}
\item \textbf{S14} \bibentry{robinson_participatory_2022}
\item \textbf{S15} \bibentry{bella_challenges_2014}
\item \textbf{S16} \bibentry{koshima_model-driven_2016}
\item \textbf{S17} \bibentry{goonetilleke_enhancing_2020}
\item \textbf{S18} \bibentry{cleland_contextualizing_2015}
\item \textbf{S19} \bibentry{beltran_smart_2022}
\item \textbf{S20} \bibentry{faria_active_2022}
\item \textbf{S21} \bibentry{le_design_2014}
\item \textbf{S22} \bibentry{maria_luisa_rodriguez-almendros_design_2021}
\item \textbf{S23} \bibentry{sokoler_embracing_2007}
\item \textbf{S24} \bibentry{hazwani_mohd_mohadis_designing_2016}
\item \textbf{S25} \bibentry{kuusik_home_2012}
\item \textbf{S26} \bibentry{casas_user_2008}
\item \textbf{S27} \bibentry{scandurra_user_2008}
\item \textbf{S28} \bibentry{berge_acceptability_2022}
\item \textbf{S29} \bibentry{willard_development_2018}
\item \textbf{S30} \bibentry{pal_internet--things_2018}
\item \textbf{S31} \bibentry{pal_embracing_2019}
\item \textbf{S32} \bibentry{ziefle_results_2020}
\item \textbf{S33} \bibentry{macdonald_designing_2007}
\item \textbf{S34} \bibentry{umm_e_mariya_shah_usability_2022}
\item \textbf{S35} \bibentry{schafer_survey-based_2019}
\item \textbf{S36} \bibentry{van_t_klooster_virtual_2011}
\item \textbf{S37} \bibentry{meiland_participation_2014}
\item \textbf{S38} \bibentry{tran-nguyen_mobile_2022}
\item \textbf{S39} \bibentry{pater_addressing_2017}
\item \textbf{S40} \bibentry{van_hoof_what_2016}
\item \textbf{S41} \bibentry{curumsing_emotion-oriented_2019}
\item \textbf{S42} \bibentry{ariaeinejad_user-centered_2016}
\item \textbf{S43} \bibentry{kawamoto_visuospatial_2017}
\item \textbf{S44} \bibentry{stamm_exergames_2019}
\item \textbf{S45} \bibentry{gollasch_age-related_2021}
\item \textbf{S46} \bibentry{ferreira_elderly_2014}
\item \textbf{S47} \bibentry{breidenbach_development_2022}
\item \textbf{S48} \bibentry{mohamad_key_2022}
\item \textbf{S49} \bibentry{martin-hammond_engaging_2018}
\item \textbf{S50} \bibentry{davidson_what_2013}
\item \textbf{S51} \bibentry{lee_steps_2017}
\item \textbf{S52} \bibentry{jorge_adaptive_2001}
\item \textbf{S53} \bibentry{sorgalla_improving_2017}
\item \textbf{S54} \bibentry{vasconcelos_designing_2012}
\item \textbf{S55} \bibentry{costa_designing_2021}
\item \textbf{S56} \bibentry{aziz_requirement_2016}
\item \textbf{S57} \bibentry{askenas_supporting_2020}
\item \textbf{S58} \bibentry{lindsay_empathy_2012}
\item \textbf{S59} \bibentry{verhoeven_mobiles_2016}
\item \textbf{S60} \bibentry{wang_mobile_2016}
\item \textbf{S61} \bibentry{compagna_limits_2015}
\item \textbf{S62} \bibentry{cahill_design_2018}
\item \textbf{S63} \bibentry{chen_understanding_2021}
\item \textbf{S64} \bibentry{teixeira_design_2017}
\item \textbf{S65} \bibentry{hendriks_designing_2013}
\item \textbf{S66} \bibentry{nielsen_user-innovated_2018}
\item \textbf{S67} \bibentry{meacham_requirements_nodate}
\item \textbf{S68} \bibentry{duh_applications_2016}
\item \textbf{S69} \bibentry{eftring_designing_2016}
\end{itemize}
\end{footnotesize}
\sloppy

\end{document}